\documentclass{article}

\usepackage{arxiv}

\usepackage[utf8]{inputenc} 
\usepackage[T1]{fontenc}    
\usepackage{hyperref}       
\usepackage{url}            
\usepackage{booktabs}       
\usepackage{amsfonts}       
\usepackage{nicefrac}       
\usepackage{microtype}      
\usepackage{lipsum}
\usepackage{moreverb,url}
\newcommand{\ignore}[1]{}
\usepackage{tikz}
\usepackage{textcomp}
\usepackage{amsmath,amsfonts,amssymb}
\usepackage[T1]{fontenc}
\usepackage[labelfont=bf]{caption}
\usepackage{cite}
\usepackage[export]{adjustbox}
\usepackage{subcaption,graphicx}
\usepackage{authblk}

\definecolor{ticolor}{HTML}{00236f}

\usepackage[explicit]{titlesec}
\renewcommand{\thesubsection}{\Alph{subsection}}

\titleformat{\section}
  {\large\sffamily\bfseries}
  {\thesection.}
  {0.5em}
  {\MakeUppercase{#1}}
  []
\titleformat{name=\section,numberless}
  {\large\sffamily\bfseries}
  {}
  {0em}
  {\MakeUppercase{#1}}
  []
\titleformat{\subsection}
  {\sffamily\bfseries}
  {\thesubsection.}
  {0.5em}
  {#1}
  []
\titleformat{\subsubsection}
  {\sffamily\small\bfseries\itshape}
  {\thesubsubsection.}
  {0.5em}
  {#1}
  []

\title{\color{ticolor}Correlative infrared optical coherence tomography and hyperspectral chemical imaging}

\author[1,*]{Ivan Zorin}
\author[2]{Rong Su}
\author[1]{Bettina Heise}
\author[3]{Bernhard Lendl}
\author[1]{Markus Brandstetter}

\affil[1]{Research Center for Non-Destructive Testing GmbH, Science Park 2,
Altenberger Str.69, 4040 Linz, Austria}
\affil[2]{Manufacturing Metrology Team, Faculty of Engineering, University of Nottingham, Nottingham NG8 1BB, UK}
\affil[3]{Institute for Chemical Technologies and Analytics, TU Wien, Getreidemarkt 9, 1060 Vienna, Austria}
\affil[*]{Corresponding author: \href{mailto:ivan.zorin@recendt.at}{ivan.zorin@recendt.at}}

\begin{document}
\maketitle

\vspace{-5pt}
\begin{abstract}
\vspace{-5pt}
Optical coherence tomography (OCT) is a high-resolution three-dimensional imaging technique that enables non-destructive measurements of surface and subsurface microstructures. 
Recent developments of OCT operating in the mid-infrared (MIR) range (around 4~\textmu m) lifted fundamental scattering limitations and initiated applied material research in formerly inaccessible fields.
The MIR spectral region, however, is also of great interest for spectroscopy and hyperspectral imaging, which allow highly selective and sensitive chemical studies of materials.
In this contribution, we introduce an OCT system (dual-band, central wavelengths of 2~\textmu m and 4~\textmu m) combined with MIR spectroscopy that is implemented as a raster scanning chemical imaging modality. The fully-integrated and cost-effective optical instrument is based on a single supercontinuum laser source (emission spectrum spanning from 1.1~\textmu m to 4.4~\textmu m).
Capabilities of the \textit{in-situ} correlative measurements are experimentally demonstrated by obtaining complex multidimensional material data, comprising morphological and chemical information, from a multi-layered composite ceramic-polymer specimen. 

\end{abstract}


\section{Introduction}

Optical coherence tomography (OCT) is a well-established non-invasive measurement technique that allows to obtain and visualize three-dimensional structural information of morphologically complex samples. Operating in the visible and near-infrared (NIR) ranges, OCT-based diagnostic methods and systems have demonstrated outstanding efficiency and capabilities in various biomedical scenarios and ophthalmology~\cite{10.1117/1.2793736}.
Nevertheless, the focus of interest within the OCT community is consequentially extending, covering non-destructive testing (NDT) applications in industrial metrology and material sciences~\cite{Stifter2007,Golde2018}.
Significant advances in OCT technology have been enabled by the rapid evolution of bright and spatially coherent mid-infrared (MIR) supercontinuum light sources~\cite{app8050707} that allowed to extend the operating wavelength range and reduce scattering effects~\cite{Su:14}.
Thus, recently reported results on the expansion of OCT into the MIR~\cite{Zorin:18,Israelsen:19} spectral region verified a hypothesis of an enhanced penetration depth into turbid materials. These solutions opened up opportunities for sub-surface dimensional measurements in the fields of, for instance, additive manufacturing~\cite{Zorin_OE:20}, pharmaceutical industry~\cite{KOLLER2011142,lin_measurement_2017}, cultural heritage diagnosis and art preservation~\cite{Cheung:14,Cheung:15,Zorin:19}. 
In general, a further increase in the central wavelength causes reduction of the axial resolution that is correlated with the coherence length and, thus, proportional to the ratio of squared wavelength and spectral bandwidth. 
However, a rational extension of the bandwidth, which shortens the coherence length, is also limited in practice since this may cause a degradation of the point-spread-function due to dispersion~\cite{Drexler}.
Therefore, in contrast to the terahertz range~\cite{Kitahara2020}, the 3~\textmu m~-~7~\textmu m part of the MIR spectral region represents a reasonable compromise, in particular, for industrial applications where high spatial resolution is a necessity. 

Since their recent commercialization, MIR supercontinuum sources have been significantly optimized in terms of their noise performance, thereby attracting significant attention within the infrared (IR) spectroscopy community. Due to the broadband emission and peculiar properties of the radiation, these sources have already launched intensive applied research in the field. Through the particular advantages over conventional thermal emitters, their efficiency was demonstrated in attenuated total reflection spectroscopy~\cite{Gasser:18}, Fourier-transform infrared (FTIR) spectroscopy~\cite{Zorin_as:20}, MIR microspectroscopy\cite{Dupont:12,Borondics:18,KilgusOE:18,Petersen:18,DASA2020100163}, gas~\cite{Grassani:19}, environmental~\cite{Saleh:19}, and stand-off spectroscopic measurements~\cite{Kumar:12,KilgusApS:18}. 
Operating in the so-called functional group region, where most of the molecules display distinct absorption lines, MIR supercontinuum sources allow to considerably extend the light-matter interaction length, thereby lowering detection limits compared to state-of-the-art FTIR instruments~\cite{Mandon:08,Zorin_as:20}. 
Due to the prevailing brightness~\cite{PETERSEN2018182} and high beam quality, supercontinuum radiation enables diffraction-limited and sensitive chemical mapping realized in reflection geometry~\cite{KilgusOE:18,Dupont:12}, thus, representing a novel and attractive tool for chemical analysis. Moreover, studies of noise properties, emission parameters and extension of the broadband spectral coverage, which is far superior to quantum cascade lasers, are the subjects of ongoing research~\cite{OB133,Liu:14,Cheng:16,Wang:17,16um,Martinez:18,Genier:19,Ghosh_2019}. Therefore, stable sources covering the entire MIR spectral range are expected to become commercially available soon.

The possible combination of these two NDT techniques~\textendash~OCT and MIR spectroscopy~\textendash~in one system is challenging and of particular interest. Spectroscopic-OCT, for instance, is a well-known and advantageous solution, which allows getting depth-resolved spectral information~\cite{Morgner:00,doi:10.1080/05704928.2017.1324876}. 
The method employs a single detector array and it is based on numerical splitting of the bandwidth in a spectral-domain OCT (SD-OCT) configuration~\cite{Drexler} into several sub-bands before performing the Fourier transform of the individual interferograms. This approach is feasible, however, impractical for a system that operates in the MIR range. One of the reasons is the absence of sensitive high-resolution detector arrays (focal plane arrays)~\textendash~the crucial component of spectroscopic- and SD-OCT~\textendash~operating in this spectral range; this technical gap is considered in detail in~\cite{Zorin:18}. However, the fundamental problem is an insufficient spectral and spatial resolution due to the contradictory requirements in the spectroscopic-OCT system: a high spectral resolution requires a narrow spectral bandwidth but confines the axial resolution of SD-OCT. For example, in order to resolve CH (carbon hydrogen functional chemical group) stretching vibrations at around 3320~nm (center wavelength, corresponding to 3010~cm\textsuperscript{-1} center wavenumber), a spectral resolution of 50~nm of the spectroscopic mode (i.e. OCT sub-band after numerical splitting) would result in a spatial resolution of around 100~\textmu m in the axial direction for the regular OCT imaging mode. For state-of-the-art OCT systems with low-NA (numerical aperture) focusing optics, such a resolution is inexpedient, since a maximum imaging depth, which is determined by the confocal gate, is on millimiter scale~\cite{Drexler,Popescu2011}.

In this paper, we propose a reasonable and new solution for the combination of these techniques that allows to achieve both high morphological and chemical resolution. Since the raster scanning approach is identical for OCT and spectroscopic imaging, we implement a multimodal system integrated in a single measurement head. In contrast to Spectroscopic-OCT, the modalities use separate detection arms, therefore, the resolution matching condition described above is eliminated. The SD-OCT system uses a single supercontinuum laser source and operates in dual-band mode (reported in~\cite{Zorin_OE:20}), i.e. within the sub-bands of 1.85~\textmu m~-~2.1~\textmu m (NIROCT) and 3.7~\textmu m~-~4.2~\textmu m (MIROCT). It enables high-resolution imaging in the NIR regime and penetration-enhanced measurements in the MIR band. The MIR spectroscopy modality supplements the morphological information obtained by OCT with correlative spatially-resolved spectra that are \ignore{superimposed}integrated over depth. The operational window of the spectroscopic system is 3.1~\textmu m~-~4.4~\textmu m (corresponds to 2275~cm\textsuperscript{-1}~-~3225~cm\textsuperscript{-1}). Being presented as a hyperspectral cube, the spectroscopic measurements reveal specimen features hidden for OCT; they enable chemical differentiation of constituent materials and analysis of their distribution. The detection systems are based on low-cost pyroelectric detectors~\cite{Zorin:18}.

\vspace{20pt}

\section{Multimodal infrared OCT and spectroscopy system}

\subsection{Infrared supercontinuum source}

\begin{figure}[ht]
\centering
\includegraphics[width=0.75\linewidth]{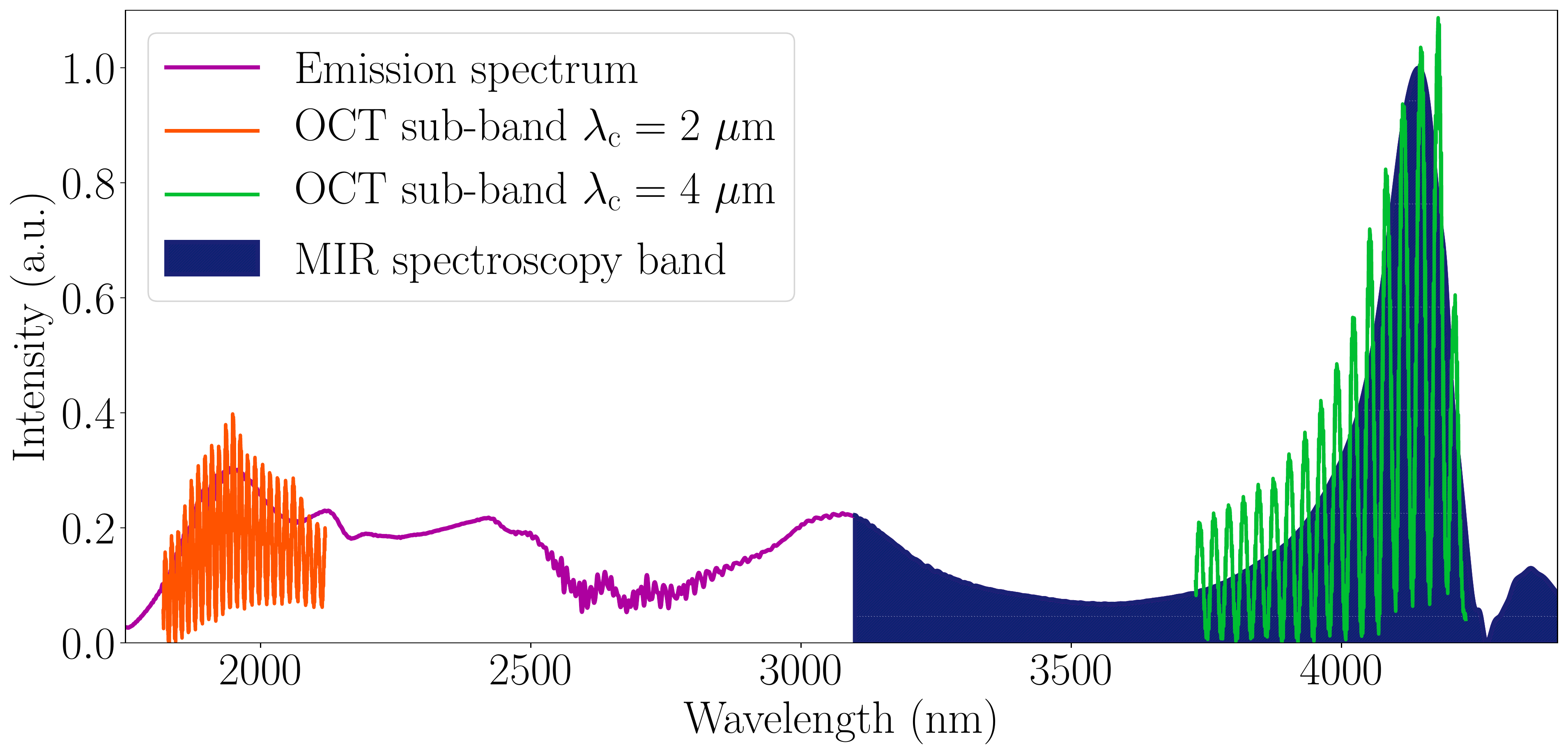}
\caption{Spectrum of the supercontinuum source measured using an FTIR spectrometer; the sensitivity curve of the MCT detector is used to correct distortions of the spectral shape; the ranges exploited for OCT and spectroscopic imaging are indicated.}
\label{fig:spectrum}
\end{figure}

The multimodal OCT and spectroscopy system reported in this study is based on a commercial pulsed supercontinuum light source from NKT Photonics (SuperK Compact Mid-IR). The supercontinua are deterministic and generated in a fluoride ZrF\textsubscript{4}-BaF\textsubscript{2}-LaF\textsubscript{3}-AlF\textsubscript{3}-NaF fiber (ZBLAN, step-index profile) pumped in the spectral band around 2050~nm (first zero dispersion wavelength) by a high-power erbium-ytterbium- and thulium-doped fiber amplifier. The broadband emission covers the range from 1.1~\textmu m to 4.4~\textmu m. Figure~\ref{fig:spectrum} shows the spectrum of the source measured using an FTIR spectrometer (Vertex 70, Bruker Optics); the bands employed for OCT and spectroscopic imaging are indicated. Edge-pass (1.65~\textmu m cut-on wavelength) and neutral density filters were used to prevent oversaturation of the Mercury Cadmium Telluride (MCT) detector. A specific spectral sensitivity of the detector was applied to reconstruct the displayed spectrum.

The beam of the laser is specified by the quality factor M\textsuperscript{2}~$\leq$~1.1; the transverse profile is the fundamental (TEM\textsubscript{00}) Gaussian mode. A parabolic mirror is used as an output collimator. The outgoing beam diameter is in the range of 3~mm~-~6~mm, measured using band-pass filters at 2~\textmu m and 4~\textmu m wavelengths correspondingly.

The repetition rate of the source is 2.5~MHz. The corresponding supercontinua pulse duration is below one nanosecond. The total average power is about 490~mW, which, in conjunction with the high quality of the beam, provides superior brightness - an essential feature for active imaging systems~\cite{svelto2010principles}. Relative pulse-to-pulse energy fluctuations were characterized using a monochromator (reported in~\cite{Zorin_as:20}); the measurements yield the mean value of around 5.8~\% over the exploited spectral range. The default operating frequency of the pyroelectric detectors depends on the sample reflectivity and magnitudes of the surface and bulk scattering and is in the range of 40~Hz~-~100~Hz. Therefore, due to the prevailing averaging, the pulse-to-pulse intensity fluctuations~\cite{Maria:17,Jensen:19} are negligible.

\subsection{Experimental system}

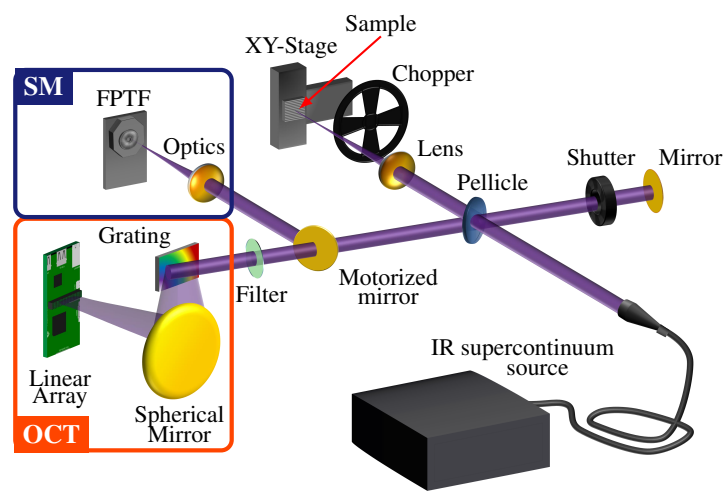
\begin{figure}[h]
\centering
\begin{tikzpicture}
 \node[anchor=south west,inner sep=0] (image) at (0,0,0) {\includegraphics[width=0.6\linewidth]{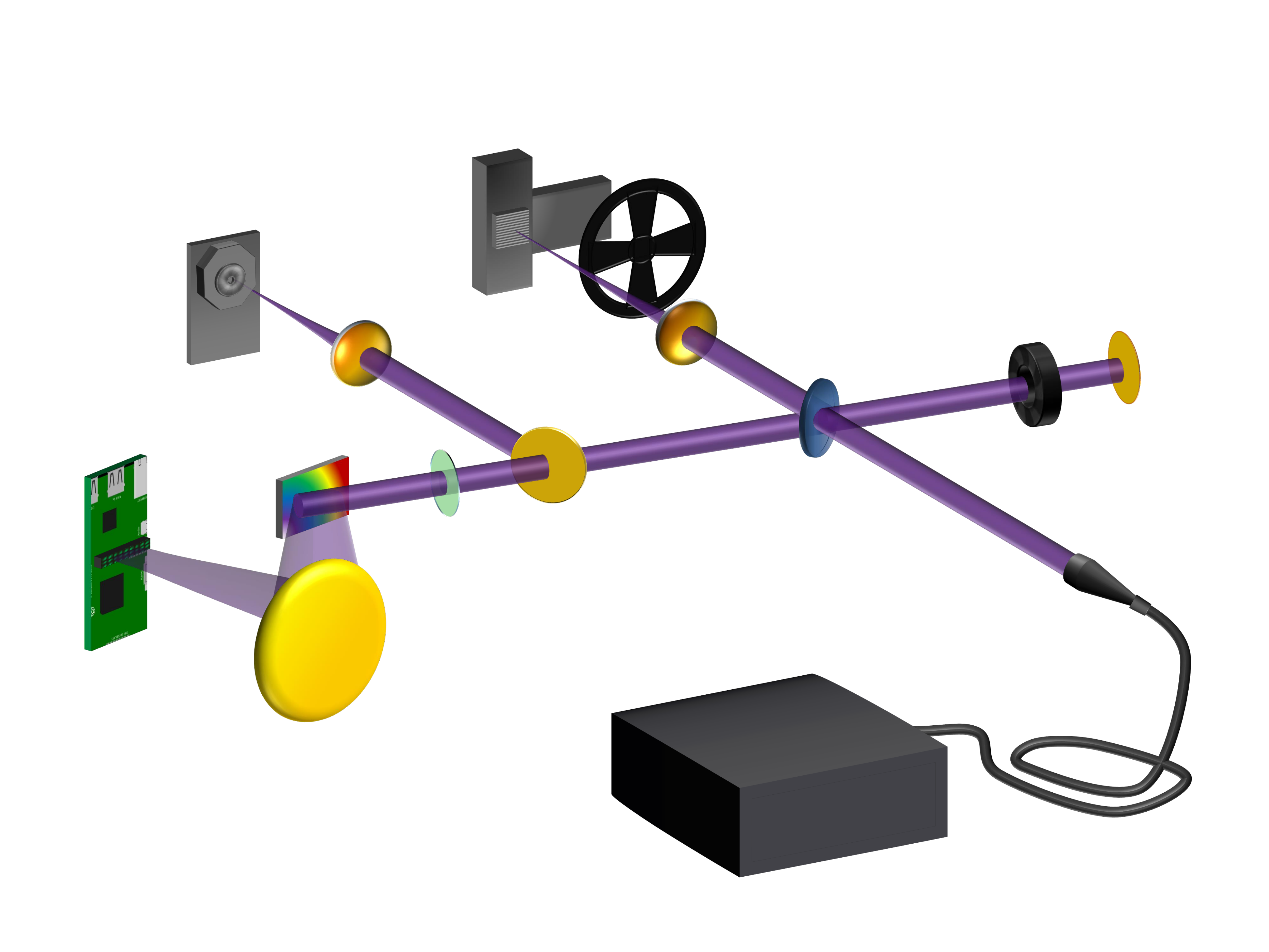}};
 \begin{scope}[x={(image.south east)},y={(image.north west)}]
 \draw (0.71,0.33) node[]{\color{black}\footnotesize IR supercontinuum};
 \draw (0.73,0.295) node[]{\color{black}\footnotesize source};
 \draw (0.59,0.82) node[]{\color{black}\footnotesize Chopper};
 \draw (0.175,0.78) node[]{\color{black}\footnotesize FPTF};
 \draw (0.09,0.28) node[]{\color{black}\footnotesize Linear};
 \draw (0.09,0.245) node[]{\color{black}\footnotesize Array};
 \draw (0.19,0.53) node[]{\color{black}\footnotesize Grating};
 \draw (0.53,0.46) node[]{\color{black}\footnotesize Motorized};
 \draw (0.53,0.425) node[]{\color{black}\footnotesize mirror};
 \draw (0.4,0.87) node[]{\color{black}\footnotesize XY-Stage};
 \draw (0.52,0.91) node[]{\color{black}\footnotesize Sample};
 \draw (0.25,0.21) node[]{\color{black}\footnotesize Spherical};
 \draw (0.25,0.18) node[]{\color{black}\footnotesize Mirror};
 \draw (0.6,0.69) node[]{\color{black}\footnotesize Lens};
 \draw (0.27,0.69) node[]{\color{black}\footnotesize Optics};
 \draw (0.67,0.63) node[]{\color{black}\footnotesize Pellicle};
 \draw (0.815,0.675) node[]{\color{black}\footnotesize Shutter};
 \draw (0.935,0.675) node[]{\color{black}\footnotesize Mirror};
 \draw (0.36,0.43) node[]{\color{black}\footnotesize Filter};
 \draw[color={rgb,255:red,255; green,69; blue,0},ultra thick,rounded corners] (0.03,0.564) rectangle (0.32,0.15);
 \draw[color={rgb,255:red,25; green,33; blue,117},ultra thick,rounded corners] (0.03,0.567) rectangle (0.32,0.8+0.03);
 \draw (0.06785,0.7711+0.03) node[fill={rgb,255:red,25; green,33; blue,117},rectangle]{\color{white}\textbf{SM}};
 \draw (0.07867,0.1787) node[fill={rgb,255:red,255; green,69; blue,0},rectangle]{\color{white}\textbf{OCT}};
 \draw [-latex,thick, color={red}] (0.52,0.89) to (0.41,0.76); 
 \end{scope}
\end{tikzpicture}
\caption{Scheme of the OCT and spectroscopic imaging system (SM), detection systems of the modalities are indicated.}
\label{fig:setup}
\end{figure}

The principal optical arrangement of the multimodal IR OCT and spectroscopy system is shown in \figurename{~\ref{fig:setup}}.
The experimental setup consists of a free-space Michelson interferometer, a measurement head (sample scanning system, focusing optics, and chopper), OCT and spectroscopy detection systems (indicated in \figurename{~\ref{fig:setup}}). The arms of the interferometer are formed by an IR pellicle beam splitter. In the reference arm, an unprotected gold mirror is used. Due to optimal dispersion and transmission profiles, a BaF\textsubscript{2} lens (50~mm focal length) is employed in the sample arm to focus the probing beam onto the specimen. The measurement head also includes a mechanical chopper to perform light intensity modulation as required for pyroelectric detectors to operate~\cite{irdet,Marshall}. The modulation system is internally synchronized with both modalities.

The detection systems of the two modalities are optically separated, therefore, the beam exiting the interferometer is reflected by a motorized mirror (synchronized with a shutter) either to a wide-band Fabry-P\'erot tunable filter spectrometer (FPTF, LFP-3144C, InfraTec GmbH) or to a grating-based OCT spectrometer. The mirror is replaceable by e.g. an IR beamsplitter, thus, allowing to perform independent measurements, but in trade-off with a magnitude of the signal. Besides, the flexibility of such a solution is limited due to the specification lag of optics and filters in this spectral range, a standard splitting ratio for the MIR beam splitters is 50:50 and invariant due to the current technology state.

The spatial resolution of the imaging system is identical for both modalities and determined by the quality of the beam and focusing optics, and, in general, is wavelength dependent. To specify the maximum and minimum resolution, a 1951 USAF resolution test target was measured at spectral extremes (around 2~\textmu m and 4~\textmu m) by means of the OCT detection modality~(see details in \hyperref[section:oct]{section D}); the results yield the spatial resolution (Rayleigh criterion) between 39.37~\textmu m (3\textsuperscript{rd} group, width of 5\textsuperscript{th} element) and 11.05~\textmu m (5\textsuperscript{th} group, width of 4\textsuperscript{th} element).

\subsection{Spectroscopy modality}
\begin{figure}[h]
\centering
\includegraphics[width=0.75\linewidth]{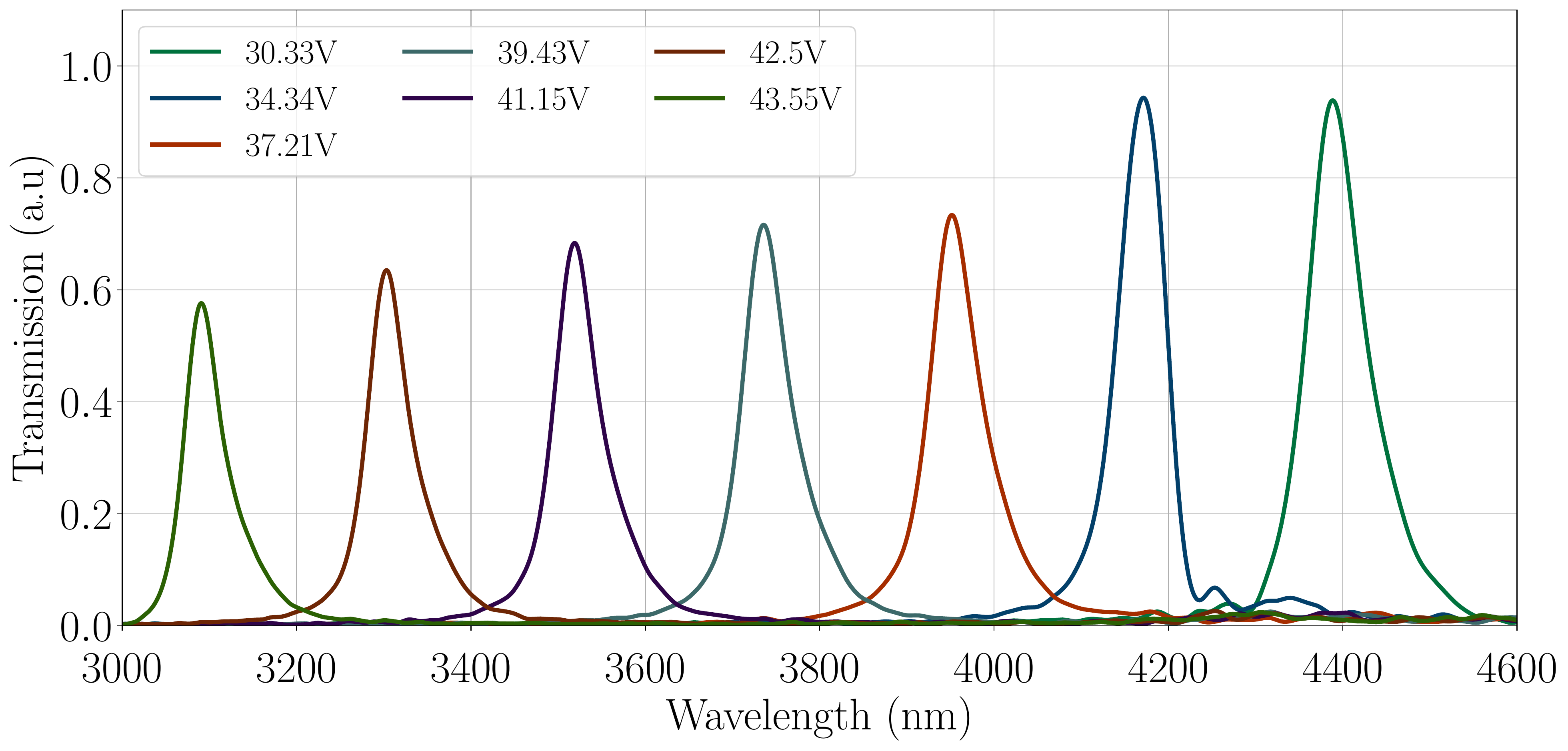}
\caption{Characterized spectral resolution (50~nm~-~75~nm at FWHM) of the Fabry-P\'erot tunable filter; transmission at different control voltages, measured using an FTIR spectrometer equipped with a thermal emitter; absorption of CO2 around 4235~nm is observed.}
\label{fig:fptf_trans}
\end{figure}

The compact and low-cost FPTF spectrometer used in the spectroscopy modality (SM) is based on a fully-integrated MEMS-based Fabry-P\'erot etalon with an electrically 
controllable cavity length. Acting as a filter, it allows to precisely tune the transmitted spectral band that satisfies the resonance condition within the wavelength range of 3.1~\textmu m~to~4.4~\textmu m (corresponds to 2275~cm\textsuperscript{-1}~-~3225~cm\textsuperscript{-1} in wavenumbers). Figure~\ref{fig:fptf_trans} shows the transmission behavior with respect to the applied voltage, recorded using an FTIR spectrometer (Vertex 70, Bruker Optics). The characterized spectral resolution is defined by the full-width at half maximum (FWHM) of the transmitted band yielding 50~nm~-~75~nm (corresponds to $\approx$~40~cm\textsuperscript{-1}).

The mechanical time constant of the filter is 2~ms~-~20~ms. A pyroelectric LiTaO\textsubscript{3} point detector (2$\times$2~mm\textsuperscript{2} area) that is located behind the tunable filter is used to record spatially resolved reflectance spectra. Thus, the chopper used in the OCT detection system switches optimally between the modes without causing delays, asynchronization or the need to deactivate the modulator when changing to another measurement cycle. The detector detectivity of 3.6$\times$10\textsuperscript{6}~$\mathrm{cm\sqrt{Hz}/W}$ is specified for 10~Hz modulation frequency at a temperature of 25 \textdegree C. 

To perform the correlative spectroscopic measurements, the reference arm of the interferometer is blocked by a shutter. Therefore, a strong reference field is excluded to protect the system from possible optical damage. The mirror directs the collected light into the FPTF-based spectrometer, where it is focused onto the point detector using suitable optics. The spectrum of the probing beam, which is reflected by the sample, is measured by spectral tuning of the filter transmission.

\subsection{Optical coherence tomography modality}
\label{section:oct}

The OCT measurement cycle is triggered, when the shutter opens and the SD-OCT modality is activated by the motorized mirror. Therefore, \textit{in-situ} structural and spectral information is obtained. A Czerny-Turner type spectrometer is used to record spectral interferograms for the OCT modality. It is based on a blazed diffraction grating (300~lines/mm) and a pyroelectric linear array (PYROSENS, DIAS Infrared GmbH); a spherical mirror (100~mm focal length) is used to focus the light onto the detector. The array has 510 rectangular shaped pixels with a dimension of 20$\times$500~\textmu m\textsuperscript{2}, optimally suited for mirror-based design with non-toroidal surfaces~\cite{Zorin:18}.

\begin{figure}[h]
\centering
\includegraphics[width=\linewidth]{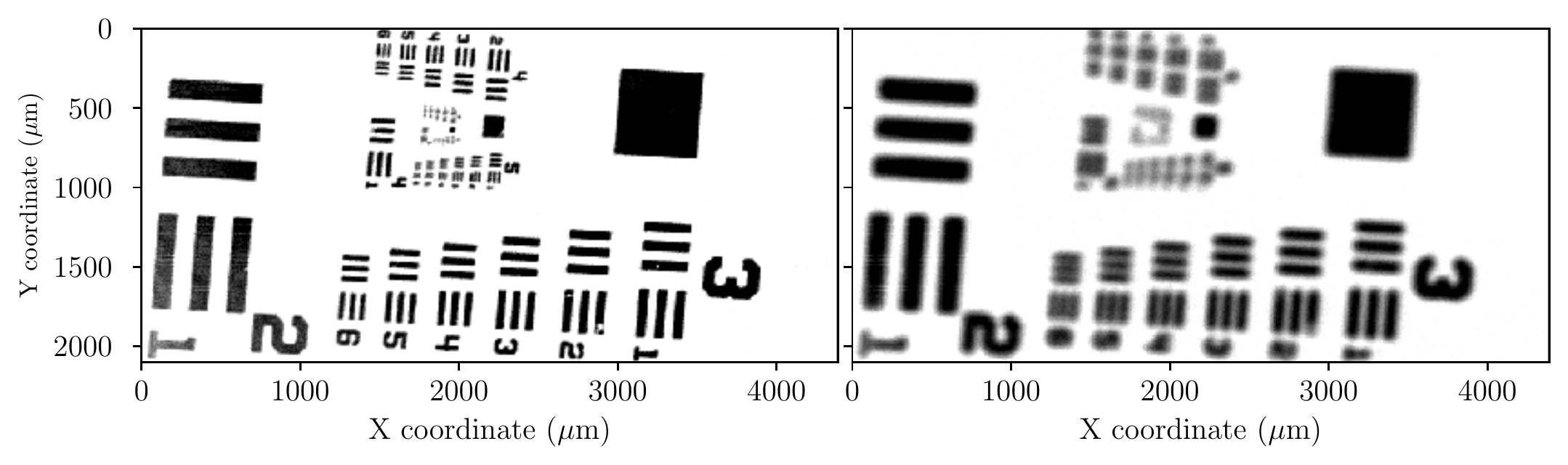}
\caption{Spatial resolution of the imaging system characterized in the extreme spectral sub-bands of the multimodal system (center wavelengths of (a)~2~\textmu m and (b)~4~\textmu m) using a 1951 USAF resolution test target; performed using OCT modality (en-face images), a neutral density filter (optical density of 0.3) was used for evaluation.}
\begin{tikzpicture}[overlay,thick]
   \draw (0,5.5) node[]{\color{black}\Large(b)};
   \draw (7.5,5.5) node[]{\color{black}\Large(a)};
 \end{tikzpicture}
\label{fig:resolution}
\end{figure}
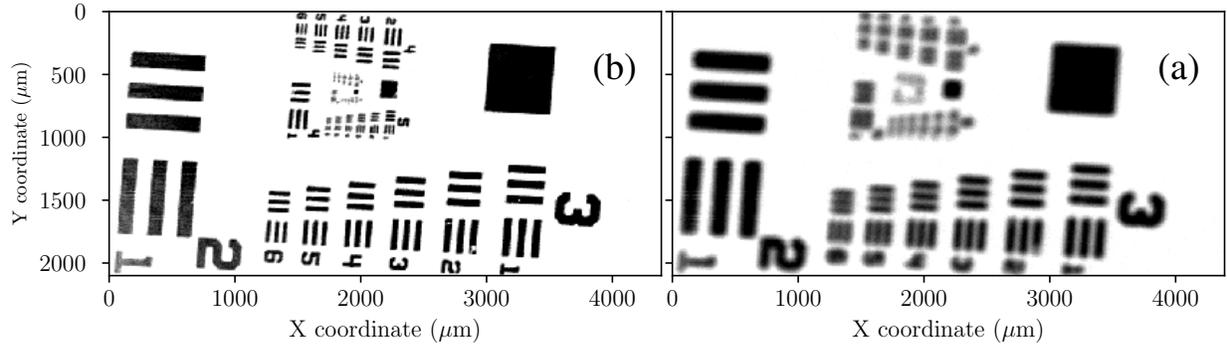

The dual-band operation mode of SD-OCT is implemented using the grating spectrometer based on a single array. Thereby, the detection of the overlapped diffraction orders of the sub-bands with doubled center wavelengths is exploited. Ghost spectra in the NIR region (1850~nm~-~2150~nm, second diffraction order) overlap with the MIR part (3750~nm~-4250~nm, first diffraction order), thus, the required band could be accessed using corresponding band-pass filters (indicated in Fig.~\ref{fig:setup}), while the second spectral region is suppressed. The axial resolution of the dual-band OCT system was characterized in~\cite{Zorin_OE:20} and is around 17~\textmu m and 37~\textmu m within the NIR and MIR ranges correspondingly; the characterization of the lateral resolution (around 11~\textmu m and 39~\textmu m respectively) is demonstrated in~\figurename{~\ref{fig:resolution}}.

The OCT A-scan rate (depth scan at a fixed lateral position) is theoretically limited by spectra acquisition speed, i.e. defined by the detection and modulation system. In practice, however, a reasonable signal-to-noise ratio is observed at chopping frequencies of 40~Hz~-~100~Hz depending on the sample reflectivity and scattering~\cite{Zorin_OE:20}.
Due to the unique features of the pyroelectric detectors, namely their sensitivity only to alternating radiation~\cite{90335,irdet}, the chopper is set in the sample arm of the interferometer. Thereby, a DC component or spectral offset usually subtracted in OCT imaging is naturally eliminated.

The post-processing algorithm employed in SD-OCT imaging is based on the method reported in~\cite{Zorin:18} and was upgraded since the reference spectrum (represents a constant signal component) is undetectable for the pyroelectric detector, thus, fewer steps are implemented. They include interpolation of the spectral interferogram to be equidistant in wavenumbers, filtering of a signal with further multiplication by a Gaussian-window. The last step involves a discrete Fourier transform of the processed interferograms for generating images in the spatial domain. A more detailed description of the principles of OCT techniques can be found in~\cite{Drexler}.

\section{Sample preparation}
In order to demonstrate the capabilities of multimodal morphological and chemical imaging, a sample based on various practically relevant industrial materials that exhibit distinct structural and spectral features was prepared. Due to the enhanced penetration of MIROCT into scattering media, a ceramic stack was selected as a basic assembly. The specimen was formed by two single-crystal Al\textsubscript{2}O\textsubscript{3} plates (0.4~mm thick), one was positioned on top of another. The top surfaces of the plates have a mean roughness 0.5~\textmu m~<~$R_a$~<~1.2~\textmu m, thus, causing strong surface scattering in the VIS and NIR ranges, where standard OCT systems are operating. Since these interfaces were positioned first to the incident light, strong specular reflection at these boundaries is reduced. The rear surfaces of the plates are epi-polished with $R_a$ < 0.2~nm.

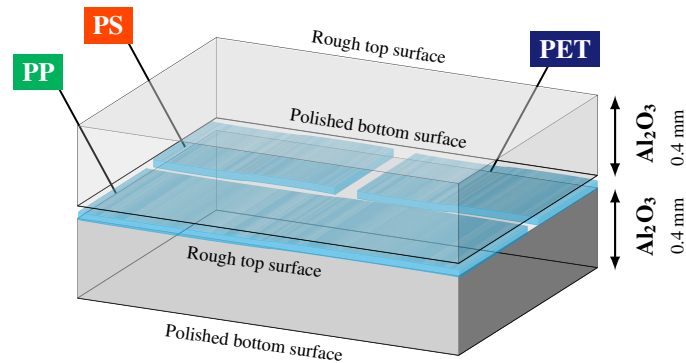
\begin{figure}[h]
\centering
\begin{tikzpicture}
 \node[anchor=south west,inner sep=0] (image) at (0,0,0) {\includegraphics[width=0.5\linewidth]{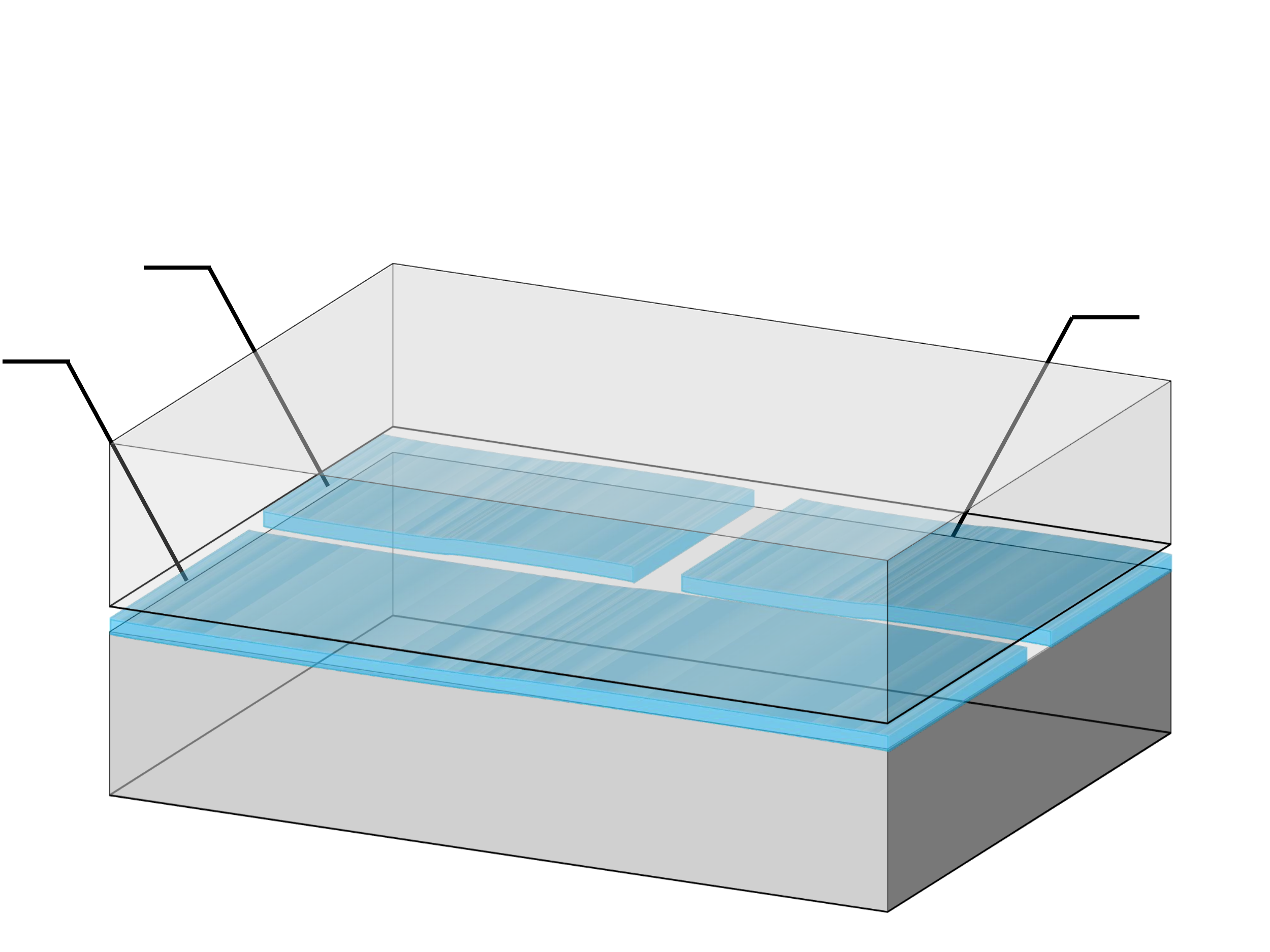}};
 \begin{scope}[x={(image.south east)},y={(image.north west)}]
 \draw [latex-latex,thick, color={black}] (0.95,0.425) to (0.95,0.6); 
 \draw [latex-latex,thick, color={black}] (0.95,0.225) to (0.95,0.4); 
  \draw (1,0.52) node[rotate=90]{\color{black}\footnotesize \textbf{Al\textsubscript{2}O\textsubscript{3}}};
 \draw (1,0.32) node[rotate=90]{\color{black}\footnotesize \textbf{Al\textsubscript{2}O\textsubscript{3}}};
  \draw (1.05,0.51) node[rotate=90]{\color{black}\scriptsize 0.4~mm};
  \draw (1.05,0.32) node[rotate=90]{\color{black}\scriptsize 0.4~mm};
 \draw (0.135,0.75) node[fill={rgb,255:red,255; green,69; blue,0},rectangle]{\color{white}\textbf{PS}};
 \draw (0.87,0.7) node[fill={rgb,255:red,25; green,33; blue,117},rectangle]{\color{white}\textbf{PET}};
 \draw (0.023,0.65) node[fill={rgb,255:red,0; green,178; blue,92},rectangle]{\color{white}\textbf{PP}};
  \draw (0.57,0.7) node[rotate=-9]{\color{black}\scriptsize Rough top surface};
  \draw (0.57,0.53) node[rotate=-9]{\color{black}\scriptsize Polished bottom surface};
 \draw (0.37,0.24) node[rotate=-8]{\color{black}\scriptsize Rough top surface};
 \draw (0.37,0.07) node[rotate=-8]{\color{black}\scriptsize Polished bottom surface};
 \end{scope}
\end{tikzpicture}
\caption{Schematic model of the multi-layered composite test sample; 50~\textmu m thick polymer films (shown in blue color)~\textendash~polyethylene terephthalate (PET), polystyrene (PS), and polypropylene (PP)~\textendash~are embedded in between the alumina ceramics; the top interfaces of the ceramic plates are rough, the bottom surfaces are epi-polished.}
\label{fig:sample}
\end{figure}

The spectral and spatial diversification of the stack was implemented employing different visible-transparent, non-scattering polymer films that were inserted in the gap between the plates. Polyethylene terephthalate (PET), polystyrene (PS), and polypropylene (PP) were selected to be accessed and measured through the turbid ceramic media. 
The films have specific spectral signatures in the exploited range: PET exhibits absorption lines induced by Phenyl-H and CH\textsubscript{2} bond stretching~\cite{doi:10.1002/pol.1958.1203312616}; for PP a slightly wider band is observed due to symmetric and asymmetric stretching vibration of CH\textsubscript{3} and CH\textsubscript{2} functional groups, and a CH shoulder-type band~\cite{Andreassen1999}\ignore{(3067~cm\textsuperscript{-1)}}. PS has distinctive and relatively strong absorption bands at higher frequencies (3082~cm\textsuperscript{-1}~-~3026~cm\textsuperscript{-1}) due to the aromatic CH stretching. Besides, asymmetric and symmetric stretching vibrations of methylene groups CH\textsubscript{2} are also observed in the absorption spectrum of PS~\cite{C4CP03516J}.
The thicknesses of the different polymer films were the same (around 50~\textmu m). Figure~\ref{fig:sample} illustrates the schematic three-dimensional model of the resulting composite sample; the constituent layers and details are indicated.

\section{Results}

\begin{figure*}
  \vspace{20pt}
\centering
 \begin{subfigure}[t]{0.36\textwidth}
  \includegraphics[width=\textwidth]{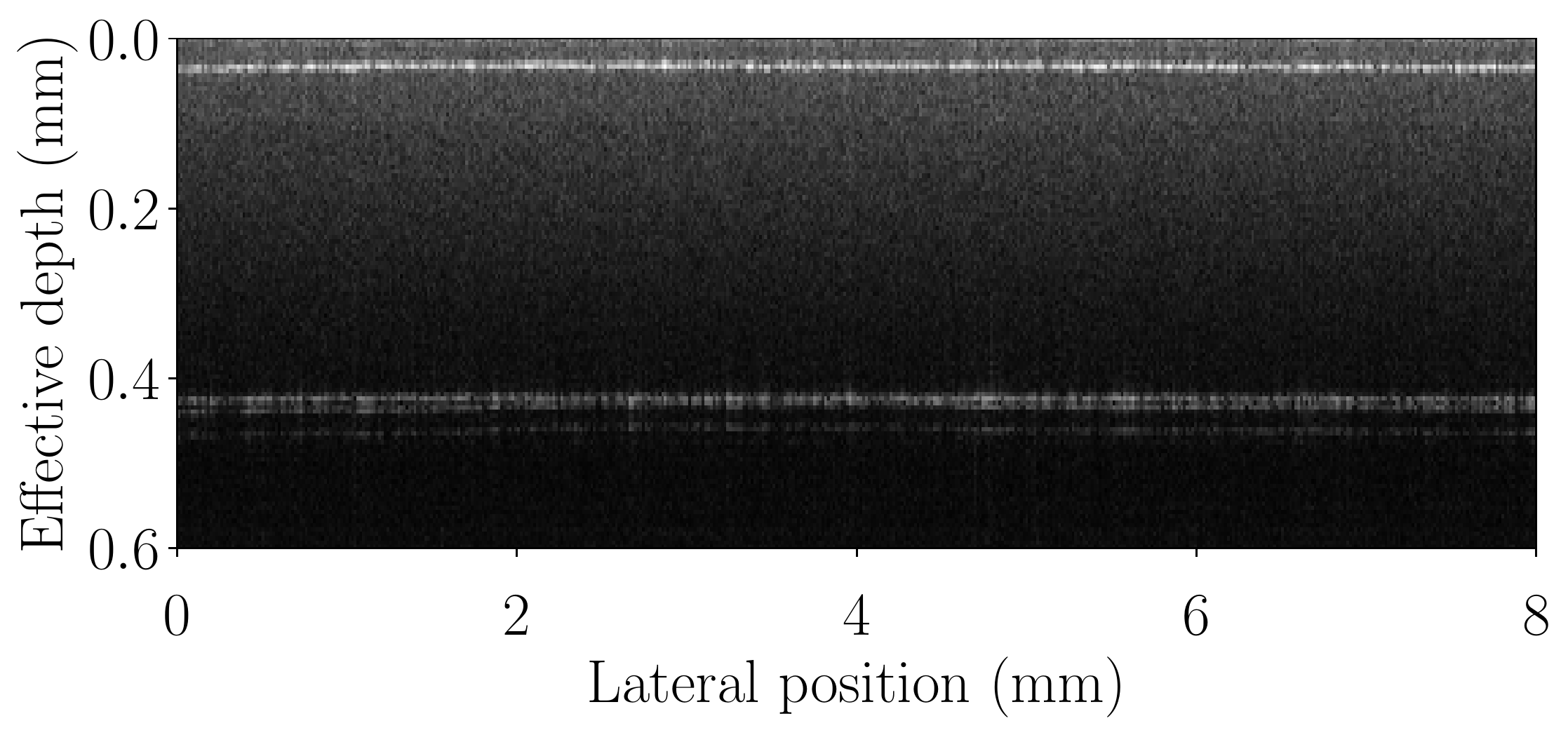}
    \vspace{20pt}
  \caption{B-scan of the top section; 2~\textmu m OCT system\label{fig:bscan_cps}}
 \end{subfigure}\hfill
 \begin{subfigure}[t]{0.36\textwidth}
  \includegraphics[width=\textwidth]{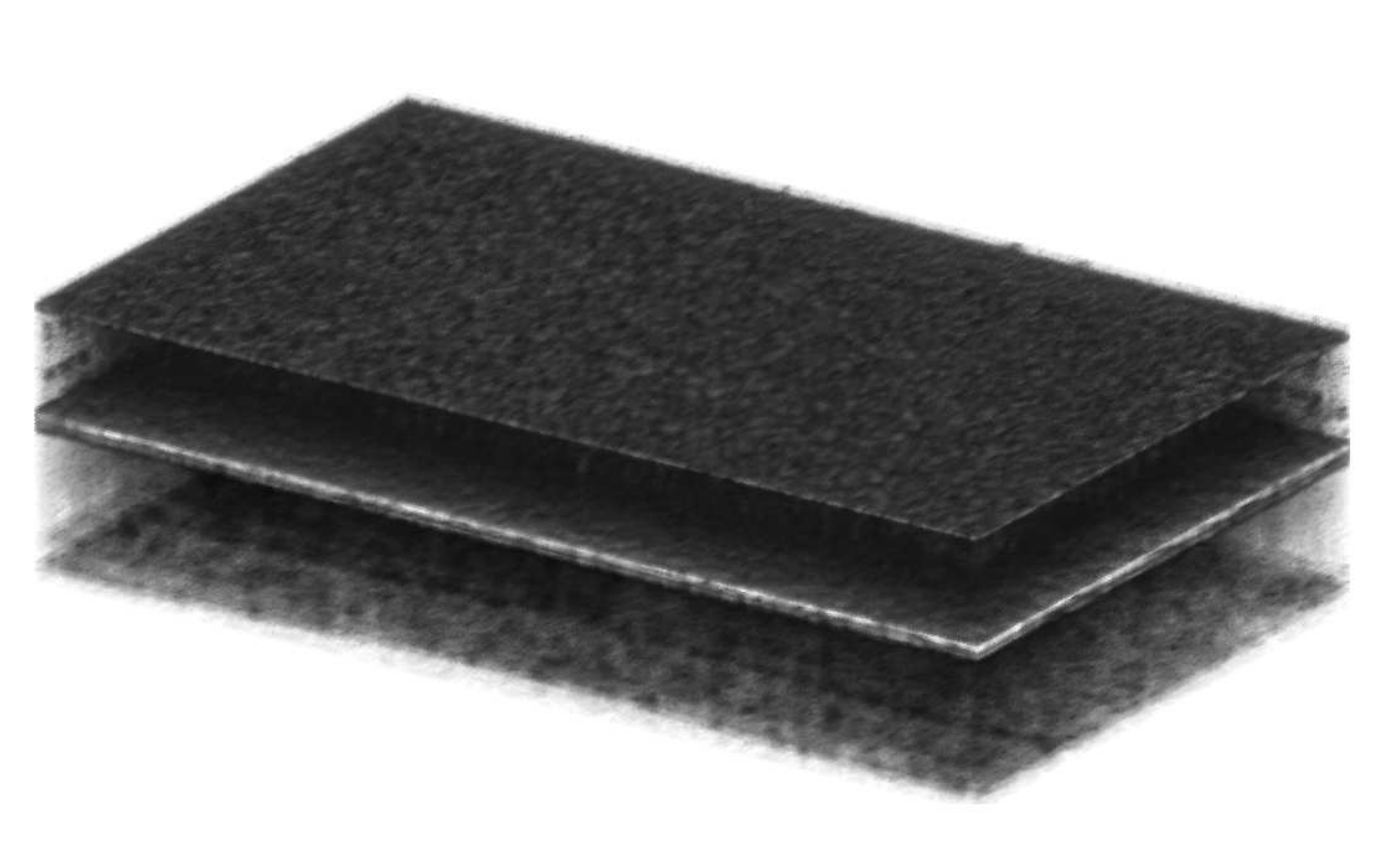}
      \vspace{20pt}
  \caption{C-scan; 4~\textmu m OCT system\label{fig:cscan}}
 \end{subfigure}\hfill
 \begin{subfigure}[t]{0.275\textwidth}
  \includegraphics[width=\textwidth]{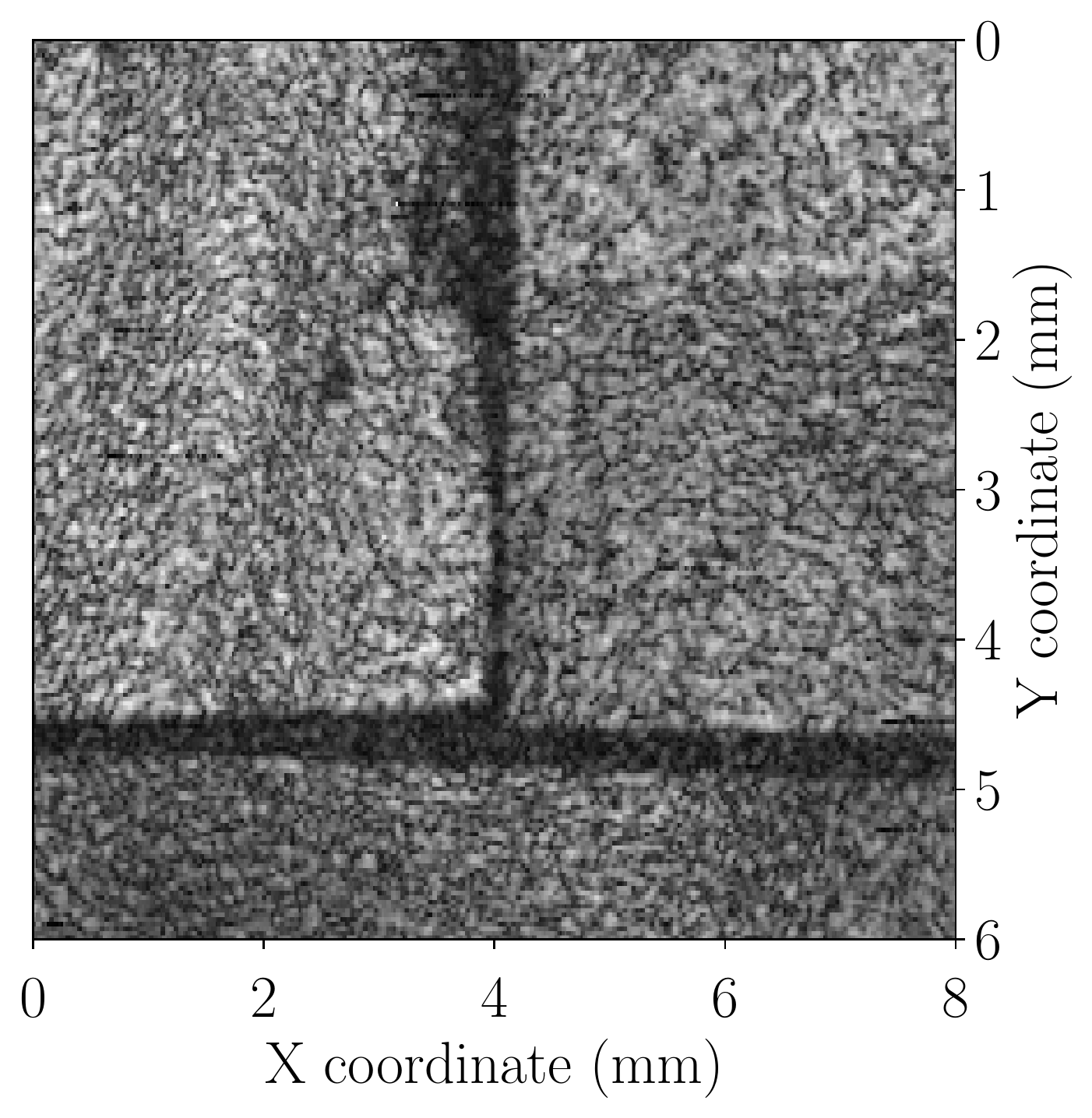}
\vspace{10pt}
  \caption{En-face image of the polymer films inserted between the plates (denoted); 4~\textmu m OCT\label{fig:enface_cps}}
 \end{subfigure}
 \caption{Tomographic imaging of the multi-layered ceramics-polymer stack by means of the dual-band IR OCT modality: (a) B-scan (high resolution 2~\textmu m OCT) of the top section of the sample, the ceramic plate and polymer films are revealed, the air gap between the ceramics and films is partially detectable (visualized and indicated in the inset); (b) C-scan of the sample obtained by means of 4~\textmu m OCT, the complete structure is accessed; (c) en-face image of the polymer films embedded in between the ceramic stack (4~\textmu m OCT) revealing morphological information but no chemical information of the polymers (types of the polymers are denoted).}
 \begin{tikzpicture}[overlay,thick]
\node[rectangle,draw={rgb,255:red,198; green,0; blue,0}, line width=1mm, inner sep=0pt] at (-4.65,8.8) {\includegraphics[scale=0.2]{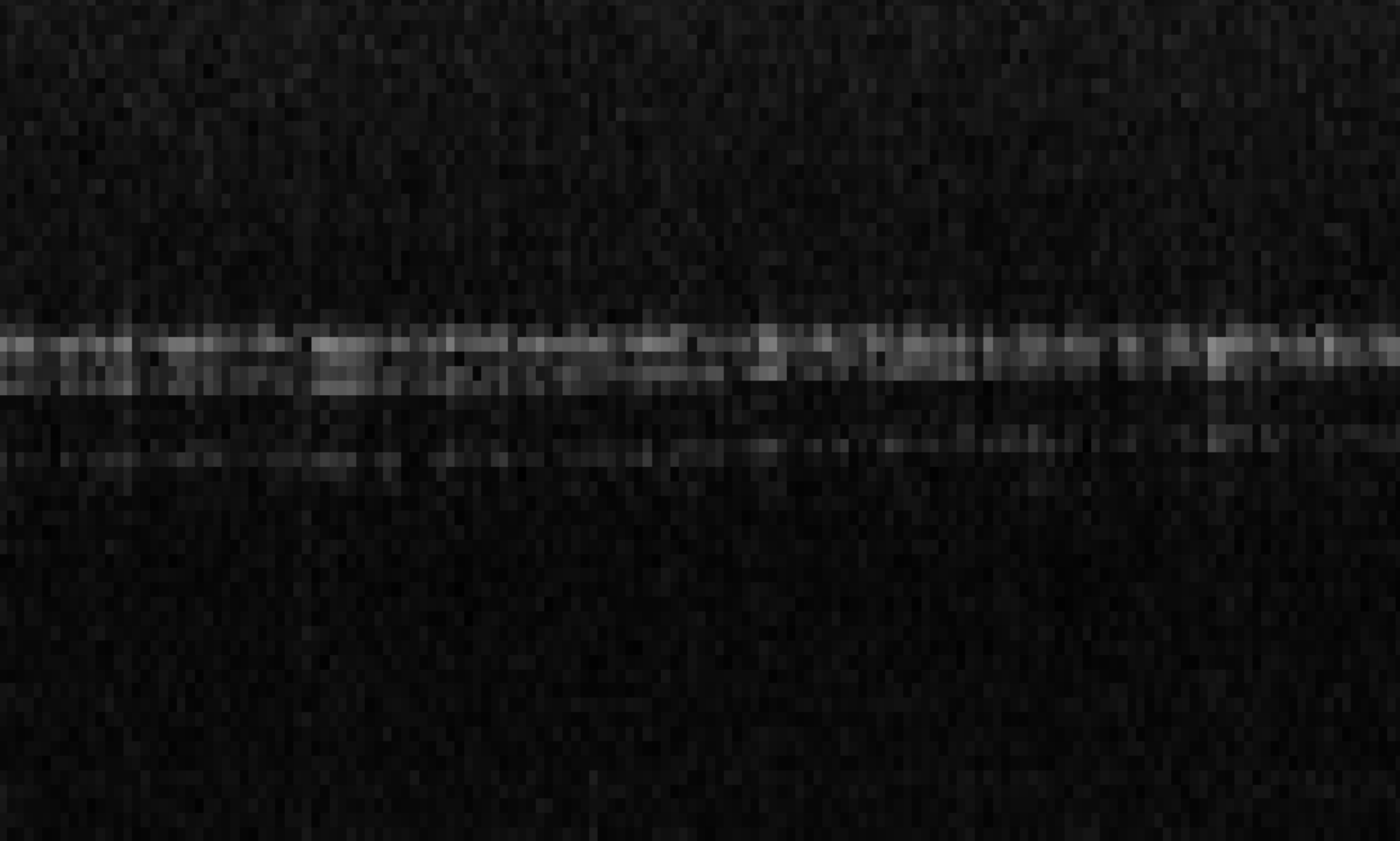}};
\draw[color={rgb,255:red,198; green,0; blue,0},ultra thick] (-8+1,5.3+0.3) rectangle (-6.3+1,6.25+0.3);
\draw [-,ultra thick, color={rgb,255:red,198; green,0; blue,0}] (-8+1,6.25+0.3) to (-6.6,7.6); 

\draw (4.4,8.9) node[fill={rgb,255:red,255; green,69; blue,0},rectangle]{\footnotesize\color{white}\textbf{PS}};
\draw (7.1,8.9) node[fill={rgb,255:red,25; green,33; blue,117},rectangle]{\footnotesize\color{white}\textbf{PET}};
\draw (5.9,5.93) node[fill={rgb,255:red,0; green,178; blue,92},rectangle]{\footnotesize\color{white}\textbf{PP}};
\draw [latex-latex,thick, color={rgb,255:red,255; green,69; blue,0}] (-2.15,7.4) to (-2.15,6.75); 
\draw [latex-latex,thick, color={rgb,255:red,255; green,69; blue,0}] (-2.15,6.7) to (-2.15,6.05); 
 \draw (-1.6,6.97) node[rotate=-15]{\color{rgb,255:red,255; green,69; blue,0}\tiny \textbf{0.4 mm}};
\draw (-1.6,6.4) node[rotate=-15]{\color{rgb,255:red,255; green,69; blue,0}\tiny \textbf{0.4 mm}};
\draw [-latex,thick, color={rgb,255:red,255; green,69; blue,0}] (-1+0.2,4.5+0.25) to (-1+0.2,6.6+0.2); 
\draw [-latex,thick, color={rgb,255:red,255; green,69; blue,0}] (-1+0.2,4.5+0.25) to (-0.25+0.2,5.75+0.25); 
\draw (-1+0.2,4.5+0.25) node[fill={rgb,255:red,255; green,69; blue,0},rectangle,rotate=0]{\color{white}\scriptsize \textbf{Al\textsubscript{2}O\textsubscript{3}}};

 \draw [dashed,ultra thick, red] (2.2-0.25,5.6+0.2) to (4-0.34,6.5+0.2); 
\draw [dashed,ultra thick, red] (2.2-0.25,5.5+0.2) to (4-0.34,6.4+0.2); 
\draw [-,ultra thick, red] (4-0.34,6.5+0.2) to (3.88,9.345); 
\draw [-,ultra thick, red] (2.2-0.25,5.5+0.2) to (3.88,5.515); 
  \draw [-,ultra thick, color={rgb,255:red,198; green,0; blue,0}] (-0.9,7.89) to (3.135,6.818); 
  \draw (2-0.3,6.92+0.07) node[fill={rgb,255:red,198; green,0; blue,0},rectangle,rotate=-14.3]{\color{white}\tiny\textbf{B-scan position}};
\draw [-latex,ultra thick, color={rgb,255:red,0; green,158; blue,69}] (-1.5,8.5) to (-1,7.5);
\draw (-1.5,8.5) node[fill={rgb,255:red,0; green,158; blue,69},rectangle]{\color{white}\scriptsize\textbf{Rough top surface}};
\draw [-latex,ultra thick, color={rgb,255:red,255; green,217; blue,0}] (-4.8,9.4) to (-4.8,8.9); 
\draw [latex-,ultra thick, color={rgb,255:red,255; green,217; blue,0}] (-4.8,8.6) to (-4.8,8.1); 
 
  \draw [-latex,ultra thick, color={white}] (-5.3,9.53) to (-5.3,9.03); 
  \draw [-latex,ultra thick, color={white}] (-5.3,8.34) to (-5.3,8.84); 
  \draw (-6,9.25) node[]{\color{white}\scriptsize\textbf{Air gap}};

 \draw (-3.8,8.3) node[]{\color{rgb,255:red,255; green,217; blue,0}\scriptsize\textbf{Polymer films}};
 \end{tikzpicture}
 \label{fig:oct_meas_cps}
\end{figure*}
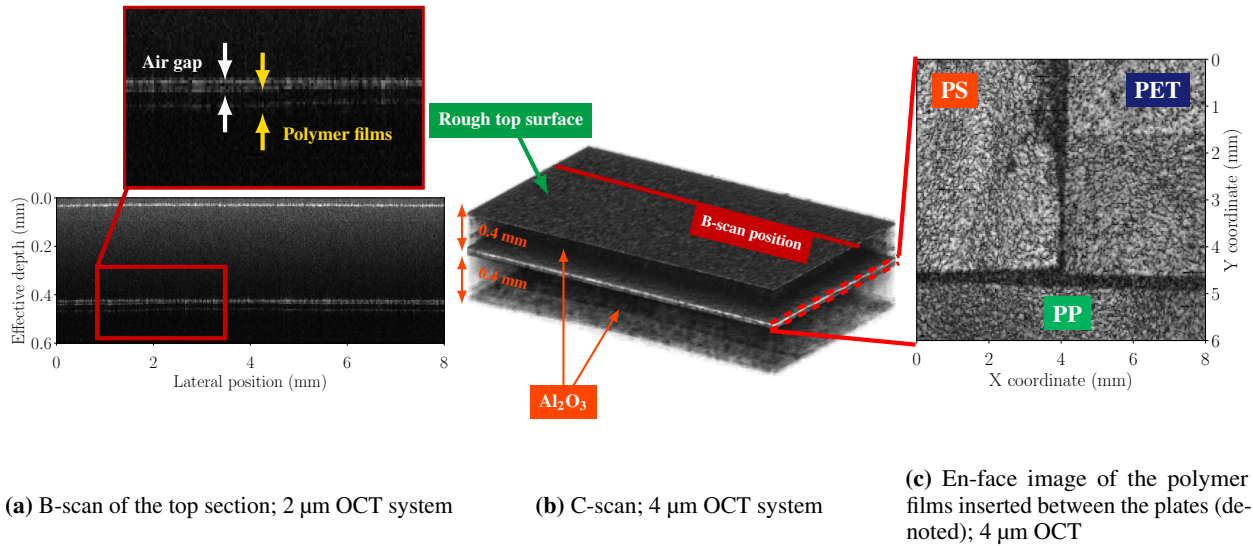

In order to demonstrate the performance of the system, correlative NIROCT, MIROCT and MIR spectroscopic measurements (area scan, 8$\times$6~mm\textsuperscript{2}) of the ceramic-polymer stack were carried out. The experimental results allow to visualize the structure and enable \textit{in-situ} identification of the embedded polymers.

The dual-band OCT measurements - depicted in \figurename{~\ref{fig:oct_meas_cps}} - reveal the entire three-dimensional volume of the multi-layered sample. The top surfaces of the plates cause no image artefacts due to the high roughness and absence of specular reflection.
The polymer films that are positioned at a depth of around 400~\textmu m are effectively detected and can be displayed using an en-face MIROCT image as shown in~\figurename{~\ref{fig:enface_cps}}. The thicknesses of the plates and films can be calculated for defined refractive indexes using high-resolution NIROCT cross-sectional images (B-scans); a typical B-scan is presented in~\figurename{~\ref{fig:bscan_cps}}. 

The air gap between the rear polished surface of the ceramic plate and the polymer films is on the same scale as the axial resolution and can be detected depending on the actual topology (i.e. at certain positions where it can be resolved). The rough top interface of the bottom plate is not detectable for NIROCT due to strong surface scattering.

Nevertheless, the complete sample morphology, which was obtained using OCT modality, does not allow differentiation of polymers due to the equal thickness and the absence of any characteristic scattering, the magnitudes of back-reflections from the different polymer interfaces are equal.

\begin{figure}[ht]
\centering
\includegraphics[width=0.8\linewidth]{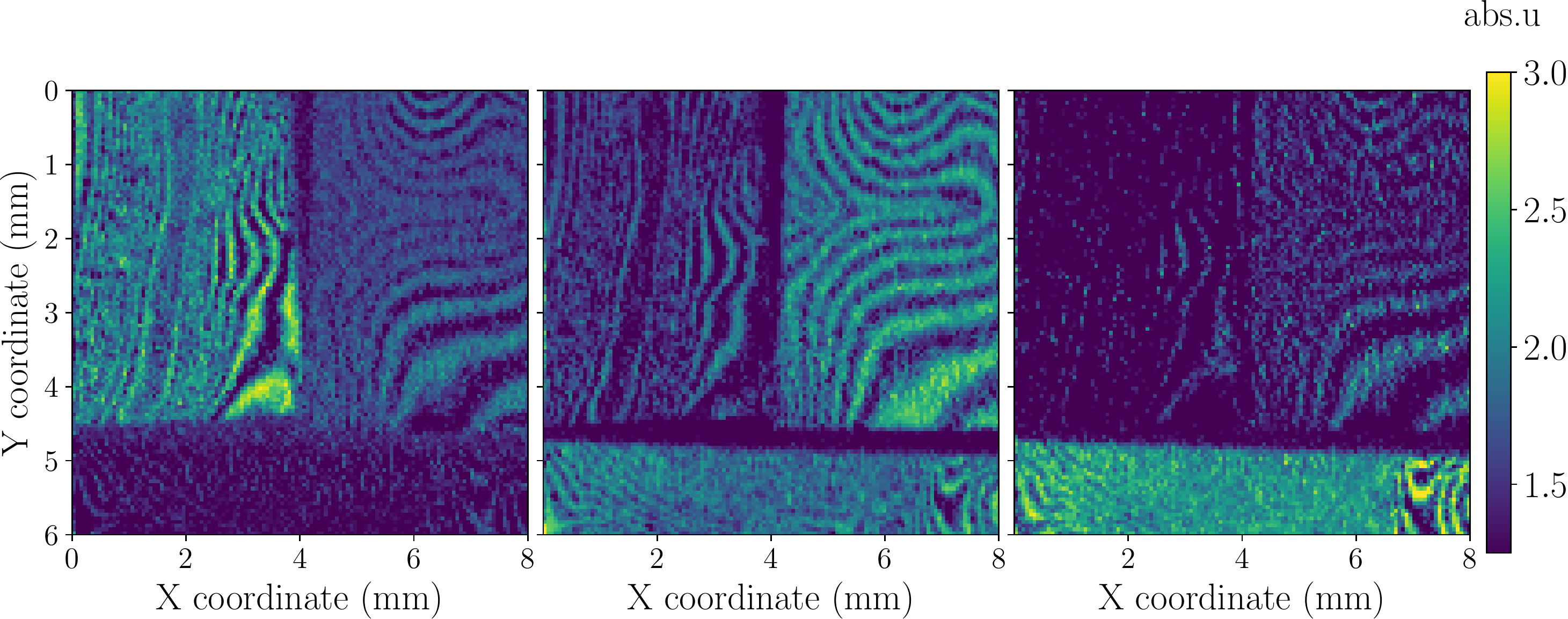}
\caption{Correlative MIR hyperspectral images of the multi-layered ceramic-polymer stack, visualized as false-color absorbance images integrated within different spectral regions (resulting in a single absorbance value for each position); the polymer films can be differentiated and identified; the clearly-visible interference patterns (fringes) are introduced by the multiple reflections within the thin air gap between the polymer films and the rear polished surface of the ceramic plate.}
\begin{tikzpicture}[overlay,thick]
  
  \draw (-4,7.2) node[]{\color{black}\footnotesize 3125 cm\textsuperscript{-1}-2976 cm\textsuperscript{-1}};
  \draw (0,7.2) node[]{\color{black}\footnotesize 2940 cm\textsuperscript{-1}-2858 cm\textsuperscript{-1}};
  \draw (4,7.2) node[]{\color{black}\footnotesize 2858 cm\textsuperscript{-1}-2702 cm\textsuperscript{-1}};
  
 \draw (-5.5,6.8-0.38) node[fill={rgb,255:red,255; green,69; blue,0},rectangle]{\color{white}\small \textbf{PS}};
 \draw (-2.7,6.8-0.38) node[fill={rgb,255:red,25; green,33; blue,117},rectangle]{\color{white}\small \textbf{PET}};
 \draw (-4,3.55) node[fill={rgb,255:red,0; green,178; blue,92},rectangle]{\color{white}\small \textbf{PP}};
 
 \draw (-1.55,6.8-0.38) node[fill={rgb,255:red,255; green,69; blue,0},rectangle]{\color{white}\small \textbf{PS}};
 \draw (1.35,6.8-0.38) node[fill={rgb,255:red,25; green,33; blue,117},rectangle]{\color{white}\small \textbf{PET}};
 \draw (0,3.55) node[fill={rgb,255:red,0; green,178; blue,92},rectangle]{\color{white}\small \textbf{PP}};
  
 \draw (2.4,6.8-0.38) node[fill={rgb,255:red,255; green,69; blue,0},rectangle]{\color{white}\small \textbf{PS}};
 \draw (5.3,6.8-0.38) node[fill={rgb,255:red,25; green,33; blue,117},rectangle]{\color{white}\small \textbf{PET}};
 \draw (4,3.55) node[fill={rgb,255:red,0; green,178; blue,92},rectangle]{\color{white}\small \textbf{PP}}; 
  
 \end{tikzpicture}
\label{fig:spect_meas_cps}
\end{figure}
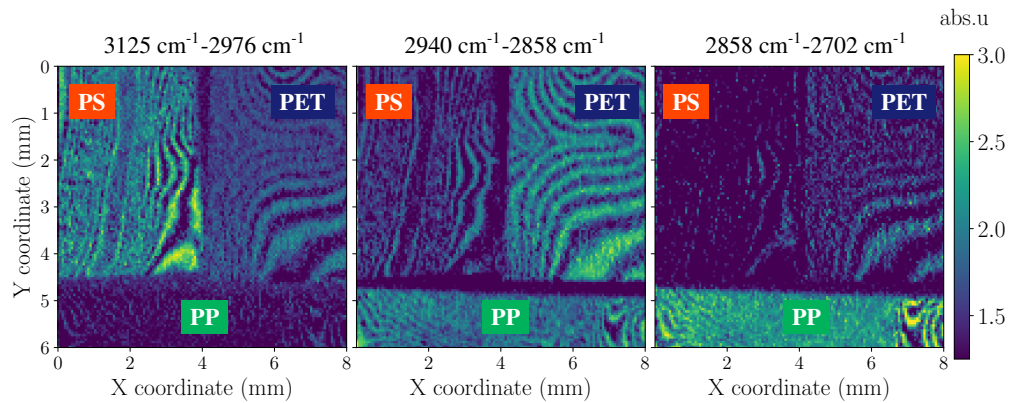

Using the MIR spectroscopy modality, spatially resolved and OCT-co-registered reflection spectra were collected, thereby a hyperspectral cube has been recorded; the acquisition time per single spectrum was around 3~sec. In order to calculate absorbance using the Beer-Lambert law~\cite{chalmers_handbook_2001}, a single reflection spectrum from the blank plates stack (no polymers inserted) was recorded as a background. In practice, the same hyperspectral data set can be used to retrieve background spectra selecting particular spatial coordinates, where e.g. no polymers are presented. As post-processing steps, a baseline correction, denoising based on principal component analysis and spectral smoothing (Savitzky-Golay filter) were applied using an open-source python-based software (Orange~\cite{doi:10.1080/08940886.2017.1338424}). 
The correlative MIR spectroscopic measurements are visualized as false-color absorbance images integrated within different spectral regions that are characteristic for chemical differentiation~(\figurename{~\ref{fig:spect_meas_cps}}). A small and varying air gap between the polymer films and rear polished surface of the plate ($R_a$ < 0.2~nm) causes interference due to multiple reflections. These effects introduce fringes in the spectroscopic images and can be also observed as an equidistant modulation in each reflection spectrum, frequencies of the modulation depend on the thickness of the air gap.

\begin{figure}
\centering
 \begin{subfigure}[t]{.35\columnwidth}
  \includegraphics[width=\columnwidth]{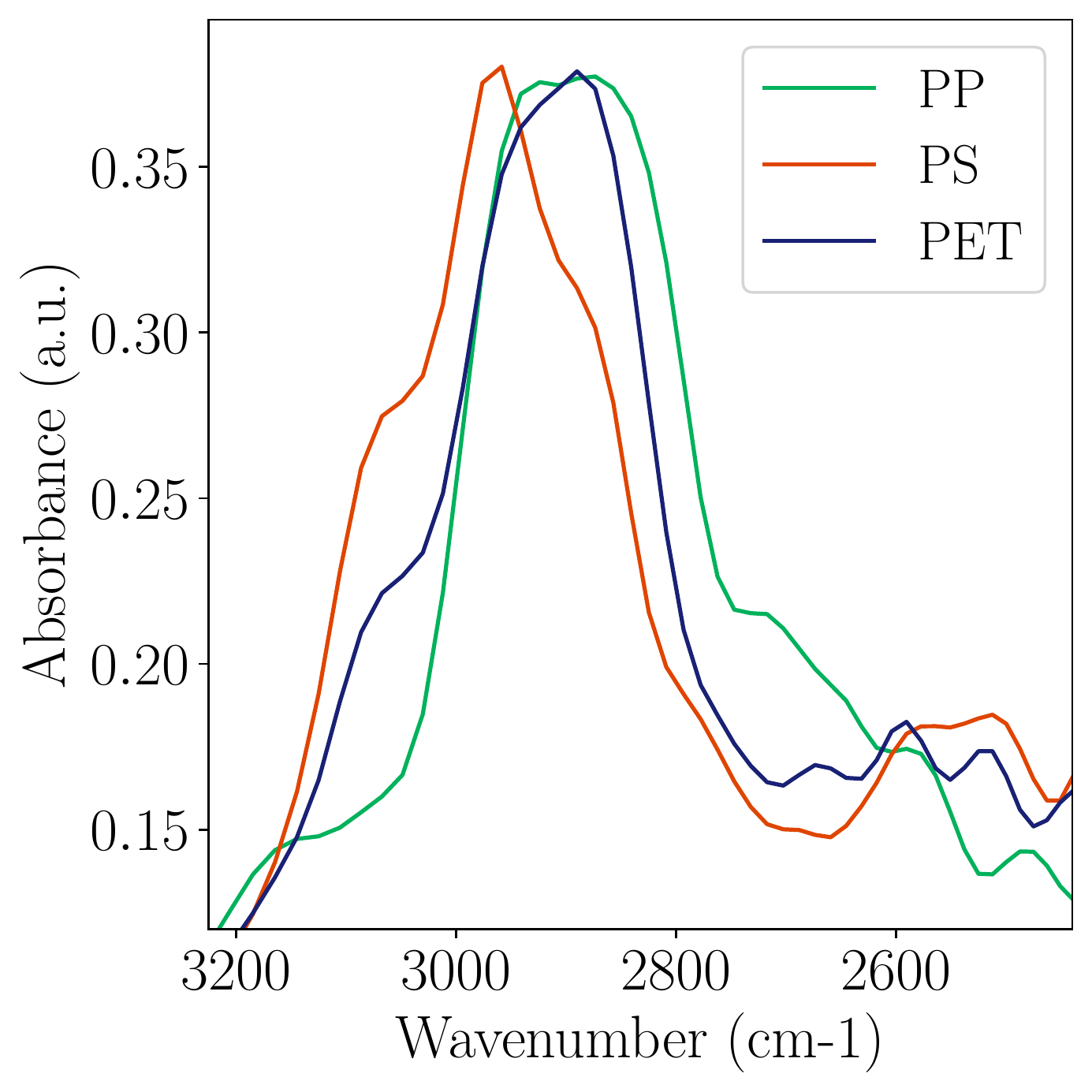}
  \caption{Absorbance spectra of the polymer films embedded in the ceramic stack, averaged over the scanned areas; experimental system, MIR spectroscopy modality (resolution 40~cm\textsuperscript{-1})\label{fig:spectra_films}}
 \end{subfigure}\hspace{10pt}
 \begin{subfigure}[t]{.35\columnwidth}
  \includegraphics[width=\columnwidth]{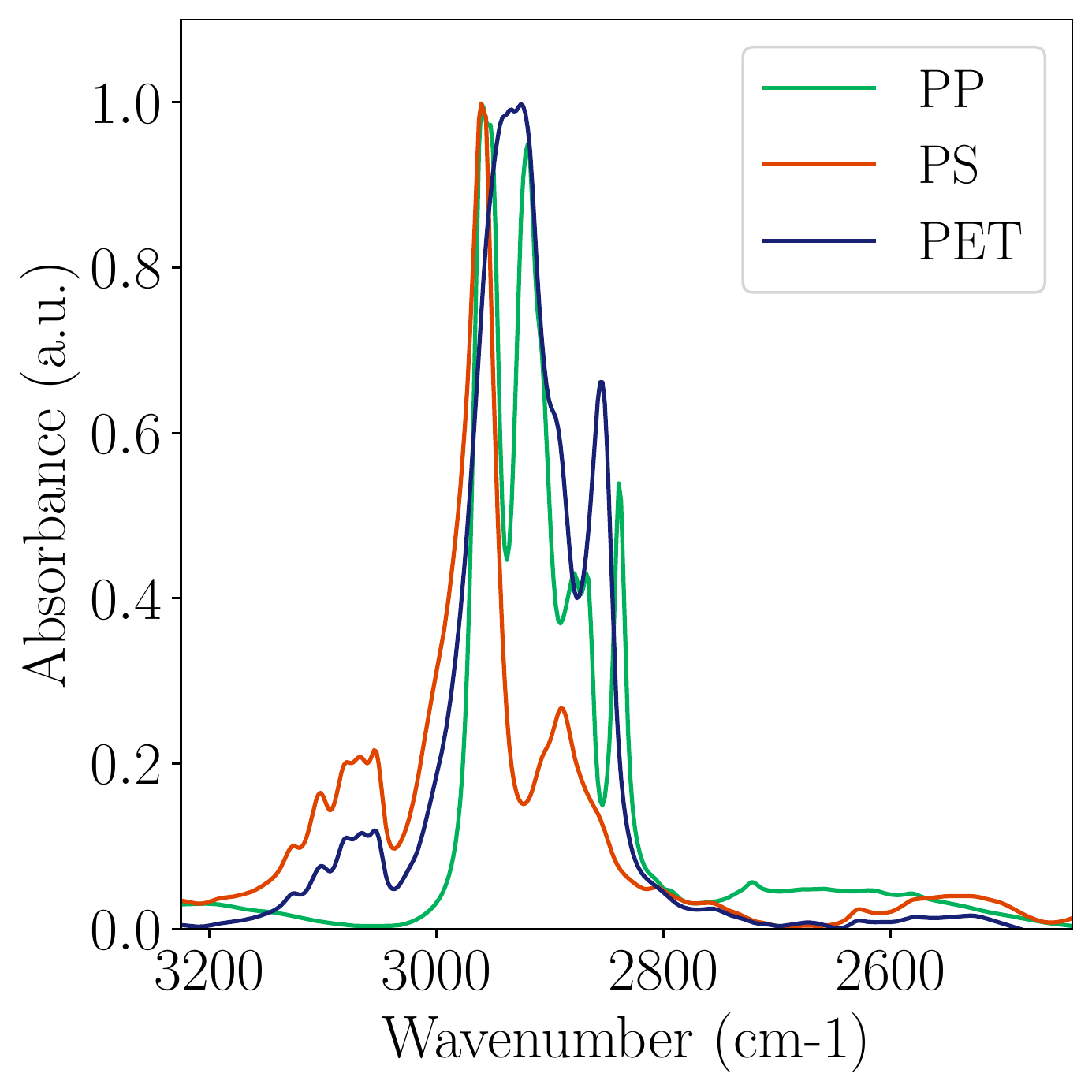}
  \caption{Reference measurements; absorbance spectra of the blank polymer films, a commercial FTIR spectrometer equipped with a thermal emitter (resolution 4~cm\textsuperscript{-1})\label{fig:films_spectra_ftir}}
 \end{subfigure}
 \caption{Spectra evaluation of the embedded polymers: (a) spatially averaged absorbance of the films and (b) reference FTIR measurements of the polymers confirming the accuracy of the spectroscopic investigation.}
 \begin{tikzpicture}[overlay,thick]
 \end{tikzpicture}
 \label{fig:spectra}
\end{figure}

As demonstrated in \figurename{~\ref{fig:spect_meas_cps}}, due to the material-dependent spectral response the evaluation of the absorbance spectra enables unambiguous differentiation of the polymers embedded in the turbid environment. Besides, each individual pixel of the recorded hyperspectral cube contains the full absorbance spectrum at the pixel’s position. Spatially averaged spectra for each polymer are shown in~\figurename{~\ref{fig:spectra}}. Effects due to thin-film interference average out in this representation. Therefore, the measurements enable identification of the polymers that display different signatures in the exploited spectral bandwidth.

In order to verify the acquired spectra, reference spectra of the polymer films were measured using a high-resolution (4~cm\textsuperscript{-1}) FTIR instrument (Lumos, Bruker Optics) that was operated in reflection mode to provide accurate data on the spectral positions of the observed absorption bands. Thus, reflectance spectra were recorded for the blank films outside the ceramic stack; back-reflected light was collected from an area of 100$\times$100~\textmu m\textsuperscript{2} and absorbance was calculated using a measured background spectrum (mirror). 
The normalized spectra~\textendash~depicted in~\figurename{~\ref{fig:films_spectra_ftir}}~\textendash~are in a good agreement with the results obtained with the developed multimodal system. Due to the lower resolution, the FPTF-based spectrometer is not able to resolve individual vibrational modes, however, the position of the bands, the envelope and trends are observed: PET and PS demonstrate absorption in the range beyond 3000~cm\textsuperscript{-1}; the width of absorption bands for PP and PET are very well consistent with FTIR measurements, especially in the range around 2800~cm\textsuperscript{-1}. 

Taking into account the spectral features and selecting a suitable spectral range to integrate absorbance (according to specific chemical signatures, as done for images in \figurename{~\ref{fig:spect_meas_cps}}), the chemical analysis of the components and their spatial distribution can be visualized for complex and spectrally diverse samples. The obtained data effectively supplement the structural information obtained by the OCT modality and reveal details that are commonly inaccessible. \ignore{Furthermore, a quantitative analysis could be implemented in order to reveal either the thickness of the media (accessible for OCT) or concentration of the particular compound.}
We would like to emphasize that the spectra were collected in reflection mode, through the 400~\textmu m thick ceramic plate with a rough surface (corresponding to 800~\textmu m, double propagation, thus, the contrast of the images can be affected. The spectral shape of the bands can also be influenced by artefacts due to interference and varying magnitude of scattering. 

Additionally, the observed interference effects (known as an autocorrelation term within OCT community~\cite{Drexler}) that, at first glance, introduce unwanted distortions in the spectra and hyperspectral images can be elegantly exploited to reconstruct the thicknesses of underlying layers using methods well-known in IR spectroscopy. Due to the significantly wider spectral bandwidth of the MIR spectroscopy modality (Fig.~\ref{fig:spectrum}), the reconstruction enables sub-OCT spatial resolution (using the standard definition for OCT~\cite{Drexler}, of approximately 5~\textmu m). Acquired reflection spectra carry distinct and singular frequencies (usually referred to by the term “fringes”) originating from constructive and destructive interference due to multiple reflections between the highly reflective polished ceramic interface (bottom surface of the top plate) and top interface of the polymers. Other possible reflections were not detected as they are significantly weaker, suppressed by e.g. strong surface scattering; the sensitivity of the SM-based reconstruction approach is weaker than that of OCT due to the absence of the strong reference field, i.e. the coherent amplification. Besides, despite the higher resolution, the maximum thickness (imaging depth) that can be accessed using this indirectly formed system of a common-path interferometer and the FPTF spectrometer (as a detector) is limited. The limitations are imposed by the lower spectral resolution of the latter, i.e. they are governed by the Nyquist-Shannon-Kotelnikov theorem; so this extension is well suited and feasible especially for measurements of relatively thin layers.

Thus, the thickness $d$ of the air gap between the top ceramic plate and the polymer films at the particular spatial coordinates can be obtained as follows~\cite{doi:10.1002/9780470106310.ch11}:
\begin{equation}
\label{eq:thickness}
d=\frac{N}{2n(\nu_1-\nu_2)},
\end{equation}
where N is a number of fringes observed in the spectral bandwidth between $\nu_1$ and $\nu_2$ (in wavenumbers), $n$ is the refractive index ($n=1$ for air) of the media between the interfaces causing interference.

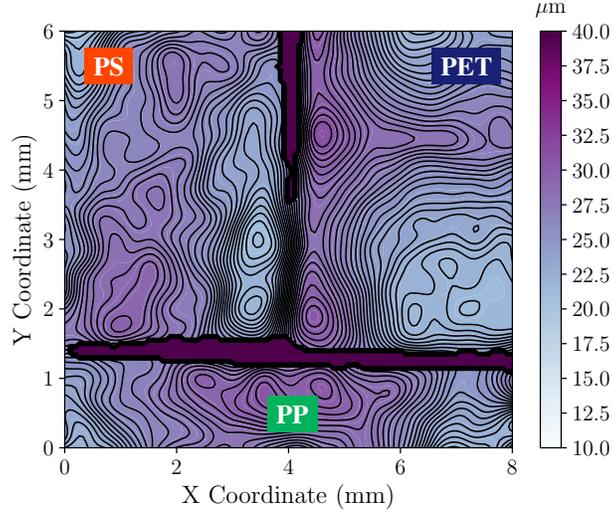
\begin{figure}[hb]
\centering
\begin{tikzpicture}
 \node[anchor=south west,inner sep=0] (image) at (0,0,0) {\includegraphics[width=0.5\linewidth]{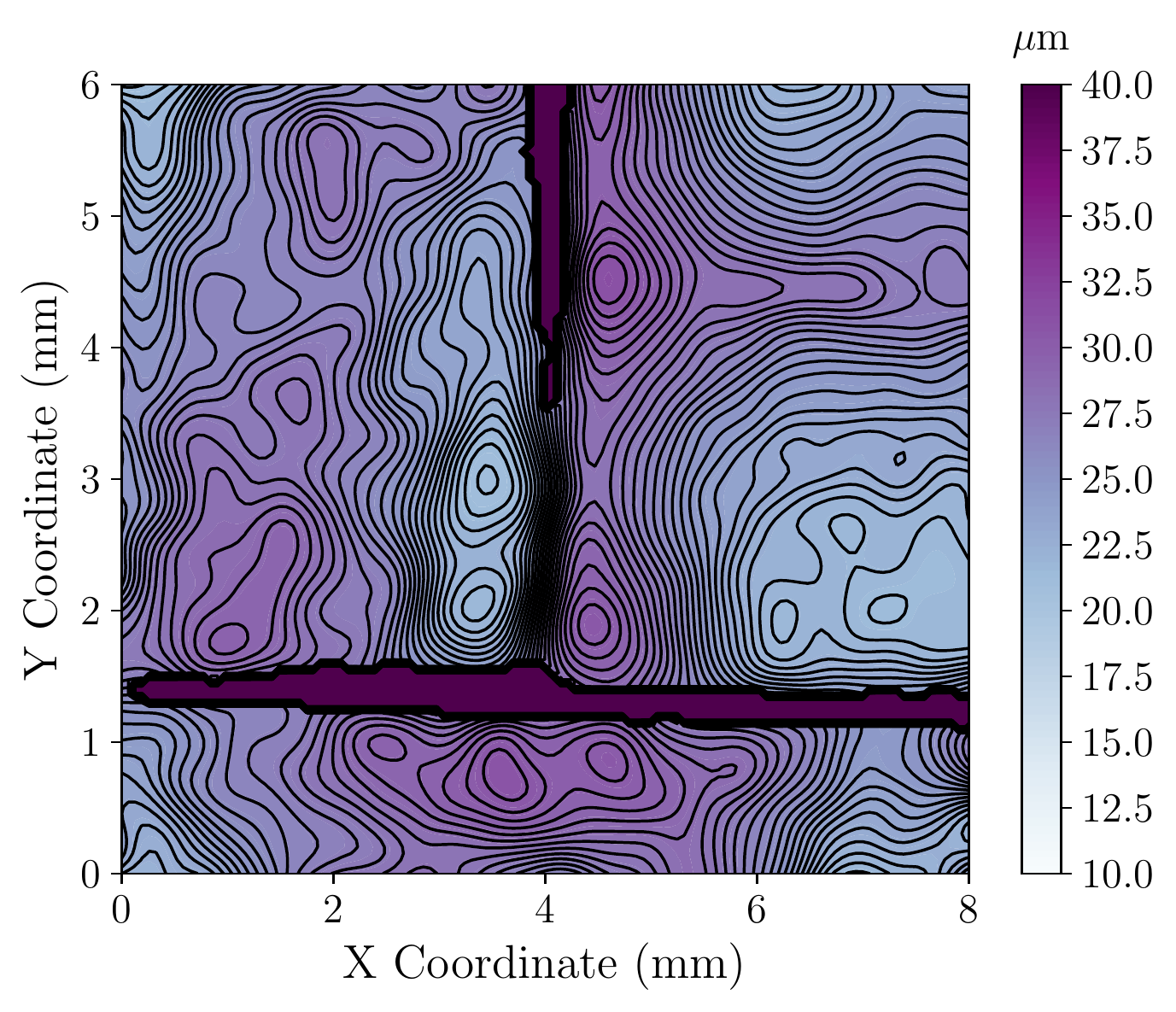}};
 \begin{scope}[x={(image.south east)},y={(image.north west)}]
 \draw (0.175,0.85) node[fill={rgb,255:red,255; green,69; blue,0},rectangle]{\color{white}\textbf{PS}};
 \draw (0.75,0.85) node[fill={rgb,255:red,25; green,33; blue,117},rectangle]{\color{white}\textbf{PET}};
 \draw (0.47,0.2) node[fill={rgb,255:red,0; green,178; blue,92},rectangle]{\color{white}\textbf{PP}};
 \end{scope}
\end{tikzpicture}
\caption{Thickness map of the air gap between the top ceramic plate and the polymer films; the profile is retrieved using the hyperspectral data, i.e. spectral fringes originated from constructive and destructive interference due to the multiple reflections within the air gap.}
\label{fig:topology}
\end{figure}

Figure~\ref{fig:topology} depicts the spatially resolved thickness map of the air gap between the top ceramic plate and the polymer films. The profile was reconstructed using~\eqref{eq:thickness}. In prior to fringes counting according to~\cite{doi:10.1002/9780470106310.ch11}, an interference-free background spectrum was subtracted from the raw spectra (for each pixel) to increase contrast and reveal pure interference patterns. Then the number of peaks was determined using a robust peak finder algorithm~\cite{lucas_hermann_negri_2017_887917}; a calculated period of the frequencies was used to access the exact fractional number of fringes. 

Thereby, the developed orthogonal modalities integrated into the single optical instrument do not only complement the data while operating with chemical and structural analysis of materials, but can successfully enhance and correct each other. The present demonstration reveals the potentials of the multimodal systems to obtain more detailed information about the samples under study~\textendash~which is of great interest for various scenarios.

\section{Conclusion and Outlook}

In this contribution, we have introduced \textit{in-situ} correlative near- and mid-infrared (NIR, MIR) optical coherence tomography (OCT) and MIR hyperspectral chemical imaging; relevant key-aspects of this unique combination were discussed in details. The developed multimodal optical system has been evaluated and experimentally demonstrated,  to the best of our knowledge, for the first time. The metrogological capabilities of the presented solution were verified by investigating a complex specimen with distinct morphology as well as chemical signatures.

The two modalities (OCT and MIR hyperspectral imaging) were united in a single measurement head and complement each other, revealing sample details that would usually be hidden for a stand-alone technique.
A single supercontinuum laser has been employed as a light source covering parts of the NIR and MIR region (1.1~\textmu m~-~4.4~\textmu m emission range).

The OCT modality operated in spectral-domain configuration (SD-OCT) exploiting a dual-band detection scheme (2~\textmu m and 4~\textmu m center wavelengths), reported in~\cite{Zorin_OE:20}, which enabled high-resolution and penetration-enhanced imaging for inspection of highly scattering samples. Using a cost-effective pyroelectric array in the MIR band, the system able to resolve sub-interfaces and detect structures embedded in a benchmark ceramic-based composite sample.

The MIR spectroscopy modality has been implemented in the attractive spectral region (3.1~\textmu m~-~4.4~\textmu m), where functional organic groups display valuable and strong absorption bands. The detection system is based on a low-cost electrically tunable Fabry-P\'erot filter equipped with a pyroelectric point detector. The spectral resolution of around 50~nm~-~75~nm ($\approx$~40~cm\textsuperscript{-1}) allowed to uniquely distinguish and chemically identify the 50~\textmu m thin polymer films (polyethylene terephthalate, polystyrene, and polypropylene) embedded in a highly scattering ceramic stack. Therefore, the spectroscopic imaging modality is particularly advantageous and feasible, as it is independent on the OCT modality; thus, no matching condition between spectral and axial resolution has to be considered.

The correlative morphological and hyperspectral imaging data set delivered by the developed system provides multidimensional information of the sample under investigation. Thereby, the emission spectrum of the single supercontinuum source is effectively and expediently exploited. Moreover, this kind of source is, in fact, an enabling technology for the presented dual-modality correlative imaging setup. Due to the current developments in the field of MIR supercontinua generation, further expansion of spectral coverage into the MIR fingerprint region, enhanced brightness, as well as decreasing prices of this novel type of laser-based light sources, are expected. Therefore, we anticipate a substantial application potential of supercontinuum laser sources for both OCT and spectroscopy in the fields of non-destructive testing, biomedical optics and beyond.

\section*{Funding}
Horizon 2020 Framework Programme (722380); Province of Upper Austria (Innovative Upper Austria 2020); EFRE Urban Innovative Actions (IWB2020); {\"O}sterreichische Forschungsf{\"o}rderungsgesellschaft (FFG) (871974).

\bibliographystyle{unsrt}  

\begin{thebibliography}{10}

\bibitem{10.1117/1.2793736}
Adam~M. Zysk, Freddy~T. Nguyen, Amy~L. Oldenburg, Daniel~L. Marks, and Stephen
  A.~Boppart M.D.
\newblock {Optical coherence tomography: a review of clinical development from
  bench to bedside}.
\newblock {\em Journal of Biomedical Optics}, 12(5):1 -- 21, 2007.

\bibitem{Stifter2007}
Stifter David.
\newblock Beyond biomedicine: a review of alternative applications and
  developments for optical coherence tomography.
\newblock {\em Applied Physics B}, 88(3):337--357, Aug 2007.

\bibitem{Golde2018}
Jonas Golde, Lars Kirsten, Christian Schnabel, Julia Walther, and Edmund Koch.
\newblock {\em Optical Coherence Tomography for NDE}, pages 1--44.
\newblock Springer International Publishing, Cham, 2018.

\bibitem{app8050707}
Shixun Dai, Yingying Wang, Xuefeng Peng, Peiqing Zhang, Xunsi Wang, and
  Yinsheng Xu.
\newblock A review of mid-infrared supercontinuum generation in chalcogenide
  glass fibers.
\newblock {\em Applied Sciences}, 8(5), 2018.

\bibitem{Su:14}
Rong Su, Mikhail Kirillin, Ernest~W. Chang, Ekaterina Sergeeva, Seok~H. Yun,
  and Lars Mattsson.
\newblock Perspectives of mid-infrared optical coherence tomography for
  inspection and micrometrology of industrial ceramics.
\newblock {\em Opt. Express}, 22(13):15804--15819, Jun 2014.

\bibitem{Zorin:18}
Ivan Zorin, Rong Su, Andrii Prylepa, Jakob Kilgus, Markus Brandstetter, and
  Bettina Heise.
\newblock Mid-infrared fourier-domain optical coherence tomography with a
  pyroelectric linear array.
\newblock {\em Opt. Express}, 26(25):33428--33439, Dec 2018.

\bibitem{Israelsen:19}
Niels~M. Israelsen, Christian~R. Petersen, Ajanta Barh, Deepak Jain, Mikkel
  Jensen, Günther Hannesschläger, Peter Tidemand-Lichtenberg, Christian
  Pedersen, Adrian Podoleanu, and Ole Bang.
\newblock Real-time high-resolution mid-infrared optical coherence tomography.
\newblock {\em Light: Science \& Applications}, 8(11):2047--7538, Jan 2019.

\bibitem{Zorin_OE:20}
Ivan Zorin, Paul Gattinger, Markus Brandstetter, and Bettina Heise.
\newblock Dual-band infrared optical coherence tomography using a single
  supercontinuum source.
\newblock {\em Opt. Express}, 28(6):7858--7874, Mar 2020.

\bibitem{KOLLER2011142}
D.M. Koller, G.~Hannesschläger, M.~Leitner, and J.G. Khinast.
\newblock Non-destructive analysis of tablet coatings with optical coherence
  tomography.
\newblock {\em European Journal of Pharmaceutical Sciences}, 44(1):142 -- 148,
  2011.

\bibitem{lin_measurement_2017}
Hungyen Lin, Yue Dong, Daniel Markl, Bryan~M. Williams, Yalin Zheng, Yaochun
  Shen, and J.~Axel Zeitler.
\newblock Measurement of the {Intertablet} {Coating} {Uniformity} of a
  {Pharmaceutical} {Pan} {Coating} {Process} {With} {Combined} {Terahertz} and
  {Optical} {Coherence} {Tomography} {In}-{Line} {Sensing}.
\newblock {\em Journal of Pharmaceutical Sciences}, 106(4):1075--1084, 04 2017.

\bibitem{Cheung:14}
C.~S. Cheung, J.~M.~O. Daniel, M.~Tokurakawa, W.~A. Clarkson, and H.~Liang.
\newblock Optical coherence tomography in the 2~\textmu m wavelength regime for
  paint and other high opacity materials.
\newblock {\em Opt. Lett.}, 39(22):6509--6512, Nov 2014.

\bibitem{Cheung:15}
C.~S. Cheung, J.~M.~O. Daniel, M.~Tokurakawa, W.~A. Clarkson, and H.~Liang.
\newblock High resolution fourier domain optical coherence tomography in the 2
  $\mu$m wavelength range using a broadband supercontinuum source.
\newblock {\em Opt. Express}, 23(3):1992--2001, Feb 2015.

\bibitem{Zorin:19}
Ivan Zorin, Jakob Kilgus, Rong Su, Bernhard Lendl, Markus Brandstetter, and
  Bettina Heise.
\newblock {Multimodal mid-infrared optical coherence tomography and
  spectroscopy for non-destructive testing and art diagnosis}.
\newblock In Haida Liang, Roger Groves, and Piotr Targowski, editors, {\em
  Optics for Arts, Architecture, and Archaeology VII}, volume 11058, pages 74
  -- 88. International Society for Optics and Photonics, SPIE, 2019.

\bibitem{Drexler}
Wolfgang Drexler and James~G. Fujimoto.
\newblock {\em Optical Coherence Tomography, Technology and Applications}.
\newblock Springer International Publishing, 2008.

\bibitem{Kitahara2020}
Hideaki Kitahara, Masahiko Tani, and Masanori Hangyo.
\newblock Frequency-domain optical coherence tomography system in the terahertz
  region.
\newblock {\em Applied Physics B}, 126(1):22, Jan 2020.

\bibitem{Gasser:18}
Christoph Gasser, Jakob Kilgus, Michael Harasek, Bernhard Lendl, and Markus
  Brandstetter.
\newblock Enhanced mid-infrared multi-bounce atr spectroscopy for online
  detection of hydrogen peroxide using a supercontinuum laser.
\newblock {\em Opt. Express}, 26(9):12169--12179, Apr 2018.

\bibitem{Zorin_as:20}
Ivan Zorin, Jakob Kilgus, Kristina Duswald, Bernhard Lendl, Bettina Heise, and
  Markus Brandstetter.
\newblock Sensitivity-enhanced fourier transform mid-infrared spectroscopy
  using a supercontinuum laser source.
\newblock {\em Applied Spectroscopy}, 74(4):485--493, 2020.
\newblock PMID: 32096412.

\bibitem{Dupont:12}
Sune Dupont, Christian Petersen, Jan Th{\o}gersen, Christian Agger, Ole Bang,
  and S{\o}ren~Rud Keiding.
\newblock Ir microscopy utilizing intense supercontinuum light source.
\newblock {\em Opt. Express}, 20(5):4887--4892, Feb 2012.

\bibitem{Borondics:18}
F.~Borondics, M.~Jossent, C.~Sandt, L.~Lavoute, D.~Gaponov, A.~Hideur,
  P.~Dumas, and S.~F\'{e}vrier.
\newblock Supercontinuum-based fourier transform infrared spectromicroscopy.
\newblock {\em Optica}, 5(4):378--381, Apr 2018.

\bibitem{KilgusOE:18}
Jakob Kilgus, Gregor Langer, Kristina Duswald, Robert Zimmerleiter, Ivan Zorin,
  Thomas Berer, and Markus Brandstetter.
\newblock Diffraction limited mid-infrared reflectance microspectroscopy with a
  supercontinuum laser.
\newblock {\em Opt. Express}, 26(23):30644--30654, Nov 2018.

\bibitem{Petersen:18}
Christian~Rosenberg Petersen, Nikola Prtljaga, Mark Farries, Jon Ward, Bruce
  Napier, Gavin~Rhys Lloyd, Jayakrupakar Nallala, Nick Stone, and Ole Bang.
\newblock Mid-infrared multispectral tissue imaging using a chalcogenide fiber
  supercontinuum source.
\newblock {\em Opt. Lett.}, 43(5):999--1002, Mar 2018.

\bibitem{DASA2020100163}
Manoj~K. Dasa, Gianni Nteroli, Patrick Bowen, Giulia Messa, Yuyang Feng,
  Christian~R. Petersen, Stella Koutsikou, Magalie Bondu, Peter~M. Moselund,
  Adrian Podoleanu, Adrian Bradu, Christos Markos, and Ole Bang.
\newblock All-fibre supercontinuum laser for in vivo multispectral
  photoacoustic microscopy of lipids in the extended near-infrared region.
\newblock {\em Photoacoustics}, 18:100163, 2020.

\bibitem{Grassani:19}
Davide Grassani, Eirini Tagkoudi, Hairun Guo, Clemens Herkommer, Fan Yang,
  Tobias~J. Kippenberg, and Camille-Sophie Br{\'e}s.
\newblock Mid infrared gas spectroscopy using efficient fiber laser driven
  photonic chip-based supercontinuum.
\newblock {\em Nature Communications}, 10(1):1553, 2019.

\bibitem{Saleh:19}
Abba Saleh, Antti Aalto, Piotr Ryczkowski, Goery Genty, and Juha Toivonen.
\newblock Short-range supercontinuum-based lidar for temperature profiling.
\newblock {\em Opt. Lett.}, 44(17):4223--4226, Sep 2019.

\bibitem{Kumar:12}
Malay Kumar, Mohammed~N. Islam, Fred~L. Terry, Michael~J. Freeman, Allan Chan,
  Manickam Neelakandan, and Tariq Manzur.
\newblock Stand-off detection of solid targets with diffuse reflection
  spectroscopy using a high-power mid-infrared supercontinuum source.
\newblock {\em Appl. Opt.}, 51(15):2794--2807, May 2012.

\bibitem{KilgusApS:18}
Jakob Kilgus, Kristina Duswald, Gregor Langer, and Markus Brandstetter.
\newblock Mid-infrared standoff spectroscopy using a supercontinuum laser with
  compact fabry-p\'erot filter spectrometers.
\newblock {\em Appl. Spectrosc.}, 72(4):634--642, Apr 2018.

\bibitem{Mandon:08}
Julien Mandon, Evgeni Sorokin, Irina~T. Sorokina, Guy Guelachvili, and Nathalie
  Picqu\'{e}.
\newblock Supercontinua for high-resolution absorption multiplex infrared
  spectroscopy.
\newblock {\em Opt. Lett.}, 33(3):285--287, Feb 2008.

\bibitem{PETERSEN2018182}
Christian~R. Petersen, Peter~M. Moselund, Laurent Huot, Lucy Hooper, and Ole
  Bang.
\newblock Towards a table-top synchrotron based on supercontinuum generation.
\newblock {\em Infrared Phys. Technol}, 91:182 -- 186, 2018.

\bibitem{OB133}
Christian~R. Petersen, Uffe M{\o}ller, Irnis Kubat, Binbin Zhou, Sune Dupont,
  Jacob Ramsay, Trevor Benson, Slawomir Sujecki, Nabil Abdel-Moneim, Zhuoqi
  Tang, David Furniss, Angela Seddon, and Ole Bang.
\newblock Mid-infrared supercontinuum covering the 1.4-13.3~\textmu m molecular
  fingerprint region using ultra-high na chalcogenide step-index fibre.
\newblock {\em Nature Photonics}, 8:830--834, Sep 2014.

\bibitem{Liu:14}
Kun Liu, Jiang Liu, Hongxing Shi, Fangzhou Tan, and Pu~Wang.
\newblock High power mid-infrared supercontinuum generation in a single-mode
  zblan fiber with up to 21.8 w average output power.
\newblock {\em Opt. Express}, 22(20):24384--24391, Oct 2014.

\bibitem{Cheng:16}
Tonglei Cheng, Kenshiro Nagasaka, Tong~Hoang Tuan, Xiaojie Xue, Morio
  Matsumoto, Hiroshige Tezuka, Takenobu Suzuki, and Yasutake Ohishi.
\newblock Mid-infrared supercontinuum generation spanning 2.0 to 15.1~\textmu m
  in a chalcogenide step-index fiber.
\newblock {\em Opt. Lett.}, 41(9):2117--2120, May 2016.

\bibitem{Wang:17}
Yingying Wang, Shixun Dai, Guangtao Li, Dong Xu, Chenyang You, Xin Han, Peiqing
  Zhang, Xunsi Wang, and Peipeng Xu.
\newblock 1.4-7.2~\textmu m broadband supercontinuum generation in an as-s
  chalcogenide tapered fiber pumped in the normal dispersion regime.
\newblock {\em Opt. Lett.}, 42(17):3458--3461, Sep 2017.

\bibitem{16um}
Zheming Zhao, Bo~Wu, Xunsi Wang, Zhanghao Pan, Zijun Liu, Peiqing Zhang, Xiang
  Shen, Qiuhua Nie, Shixun Dai, and Rongping Wang.
\newblock Mid-infrared supercontinuum covering 2.0-16~\textmu m in a low-loss
  telluride single-mode fiber.
\newblock {\em Laser \& Photonics Reviews}, 11(2):1700005, 2017.

\bibitem{Martinez:18}
Ramon~A. Martinez, Genevieve Plant, Kaiwen Guo, Brian Janiszewski, Michael~J.
  Freeman, Robert~L. Maynard, Mohammed~N. Islam, Fred~L. Terry, Oseas Alvarez,
  Francois Chenard, Robert Bedford, Ricky Gibson, and Agustin~I. Ifarraguerri.
\newblock Mid-infrared supercontinuum generation from 1.6 to >11\textmu m using
  concatenated step-index fluoride and chalcogenide fibers.
\newblock {\em Opt. Lett.}, 43(2):296--299, Jan 2018.

\bibitem{Genier:19}
Etienne Genier, Patrick Bowen, Thibaut Sylvestre, John~M. Dudley, Peter
  Moselund, and Ole Bang.
\newblock Amplitude noise and coherence degradation of femtosecond
  supercontinuum generation in all-normal-dispersion fibers.
\newblock {\em J. Opt. Soc. Am. B}, 36(2):A161--A167, Feb 2019.

\bibitem{Ghosh_2019}
A~N Ghosh, M~Meneghetti, C~R Petersen, O~Bang, L~Brilland, S~Venck, J~Troles,
  J~M Dudley, and T~Sylvestre.
\newblock Chalcogenide-glass polarization-maintaining photonic crystal fiber
  for mid-infrared supercontinuum generation.
\newblock {\em Journal of Physics: Photonics}, 1(4):044003, sep 2019.

\bibitem{Morgner:00}
U.~Morgner, W.~Drexler, F.~X. K\"{a}rtner, X.~D. Li, C.~Pitris, E.~P. Ippen,
  and J.~G. Fujimoto.
\newblock Spectroscopic optical coherence tomography.
\newblock {\em Opt. Lett.}, 25(2):111--113, Jan 2000.

\bibitem{doi:10.1080/05704928.2017.1324876}
Hyeong~Soo Nam and Hongki Yoo.
\newblock Spectroscopic optical coherence tomography: A review of concepts and
  biomedical applications.
\newblock {\em Applied Spectroscopy Reviews}, 53(2-4):91--111, 2018.

\bibitem{Popescu2011}
Dan~P. Popescu, Lin-P'ing Choo-Smith, Costel Flueraru, Youxin Mao, Shoude
  Chang, John Disano, Sherif Sherif, and Michael~G. Sowa.
\newblock Optical coherence tomography: fundamental principles, instrumental
  designs and biomedical applications.
\newblock {\em Biophysical Reviews}, 3(3):155, Aug 2011.

\bibitem{svelto2010principles}
O.~Svelto.
\newblock {\em Principles of Lasers}.
\newblock Springer US, 2010.

\bibitem{Maria:17}
Michael Maria, Ivan~Bravo Gonzalo, Thomas Feuchter, Mark Denninger, Peter~M.
  Moselund, Lasse Leick, Ole Bang, and Adrian Podoleanu.
\newblock Q-switch-pumped supercontinuum for ultra-high resolution optical
  coherence tomography.
\newblock {\em Opt. Lett.}, 42(22):4744--4747, Nov 2017.

\bibitem{Jensen:19}
Mikkel Jensen, Iv\'{a}n~Bravo Gonzalo, Rasmus~Dybbro Engelsholm, Michael Maria,
  Niels~M{\o}ller Israelsen, Adrian Podoleanu, and Ole Bang.
\newblock Noise of supercontinuum sources in spectral domain optical coherence
  tomography.
\newblock {\em J. Opt. Soc. Am. B}, 36(2):A154--A160, Feb 2019.

\bibitem{irdet}
Antonio Rogalski.
\newblock {\em Infrared Detectors 2nd Edition}.
\newblock CRC, 2011.

\bibitem{Marshall}
Donald~E. Marshall.
\newblock A review of pyroelectric detector technology.
\newblock In {\em Utilization of Infrared Detectors}, volume 0132, 1978.

\bibitem{90335}
A.~{Hossain} and M.~H. {Rashid}.
\newblock Pyroelectric detectors and their applications.
\newblock {\em IEEE Transactions on Industry Applications}, 27(5):824--829,
  Sep. 1991.

\bibitem{doi:10.1002/pol.1958.1203312616}
W.~W. Daniels and R.~E. Kitson.
\newblock Infrared spectroscopy of polyethylene terephthalate.
\newblock {\em Journal of Polymer Science}, 33(126):161--170, 1958.

\bibitem{Andreassen1999}
J.~Karger-Kocsis, editor.
\newblock {\em Infrared and Raman spectroscopy of polypropylene}, pages
  320--328.
\newblock Springer Netherlands, Dordrecht, 1999.

\bibitem{C4CP03516J}
D.~Olmos, E.~V. Martín, and J.~González-Benito.
\newblock New molecular-scale information on polystyrene dynamics in ps and
  ps–batio3 composites from ftir spectroscopy.
\newblock {\em Phys. Chem. Chem. Phys.}, 16:24339--24349, 2014.

\bibitem{chalmers_handbook_2001}
John~M. Chalmers and Peter~R. Griffiths, editors.
\newblock {\em Handbook of {Vibrational} {Spectroscopy}: {Chalmers} {Vibrat}
  {5V} {Set}}.
\newblock John Wiley \& Sons, Ltd, Chichester, UK, December 2001.

\bibitem{doi:10.1080/08940886.2017.1338424}
M.~Toplak, G.~Birarda, S.~Read, C.~Sandt, S.~M. Rosendahl, L.~Vaccari,
  J.~Demšar, and F.~Borondics.
\newblock Infrared orange: Connecting hyperspectral data with machine learning.
\newblock {\em Synchrotron Radiation News}, 30(4):40--45, 2017.

\bibitem{doi:10.1002/9780470106310.ch11}
Peter~R. Griffiths and James~A. de~Haseth.
\newblock Conventional transmission spectrometry.
\newblock In {\em Fourier Transform Infrared Spectrometry}, chapter~11, pages
  251--260. John Wiley \& Sons, Ltd, 2006.

\bibitem{lucas_hermann_negri_2017_887917}
Lucas~Hermann Negri and Christophe Vestri.
\newblock lucashn/peakutils: v1.1.0.
\newblock \url{https://doi.org/10.5281/zenodo.887917}, September 2017.

\end{thebibliography}

\end{document}